\documentclass[12pt]{article}

\usepackage{url}
\usepackage{setspace}
\usepackage{amsmath,amssymb,amscd, amsthm}
\usepackage{bbold}
\usepackage{amsfonts}
\usepackage[usenames,dvipsnames]{color}
\usepackage{graphicx}
\usepackage{subcaption}

\usepackage{bm}

\usepackage{hyperref}
\usepackage{cleveref}

\newcommand{\ZZ}{\mathbb{Z}}		
\newcommand{\CC}{\mathbb{C}}		
\newcommand{\CFT}{\mathcal{C}}
\DeclareMathOperator{\Tr}{Tr}	
\renewcommand{\Im}{\operatorname{Im}}

\newcommand{\opO}{\mathcal{O}}
\newcommand{\blockF}{\mathcal{F}}

\newcommand\rref[1]{(\ref{#1})}

\newcommand{\cali}{\includegraphics[height=4.725mm, trim = 0mm 0mm 0mm 0mm, clip]{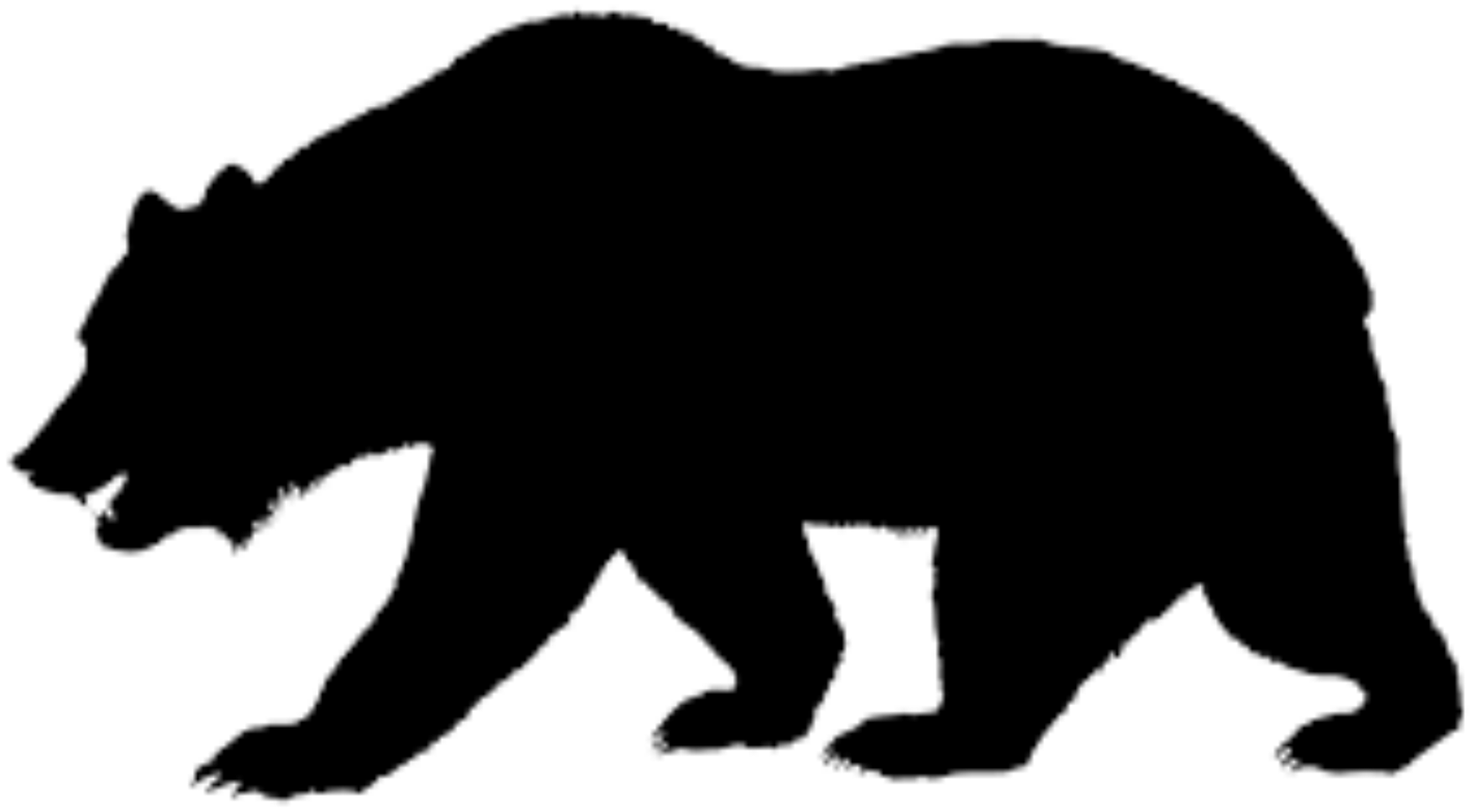}}
\newcommand{\ox}{\includegraphics[height=4.725mm, trim = 0mm 0mm 0mm 0mm, clip]{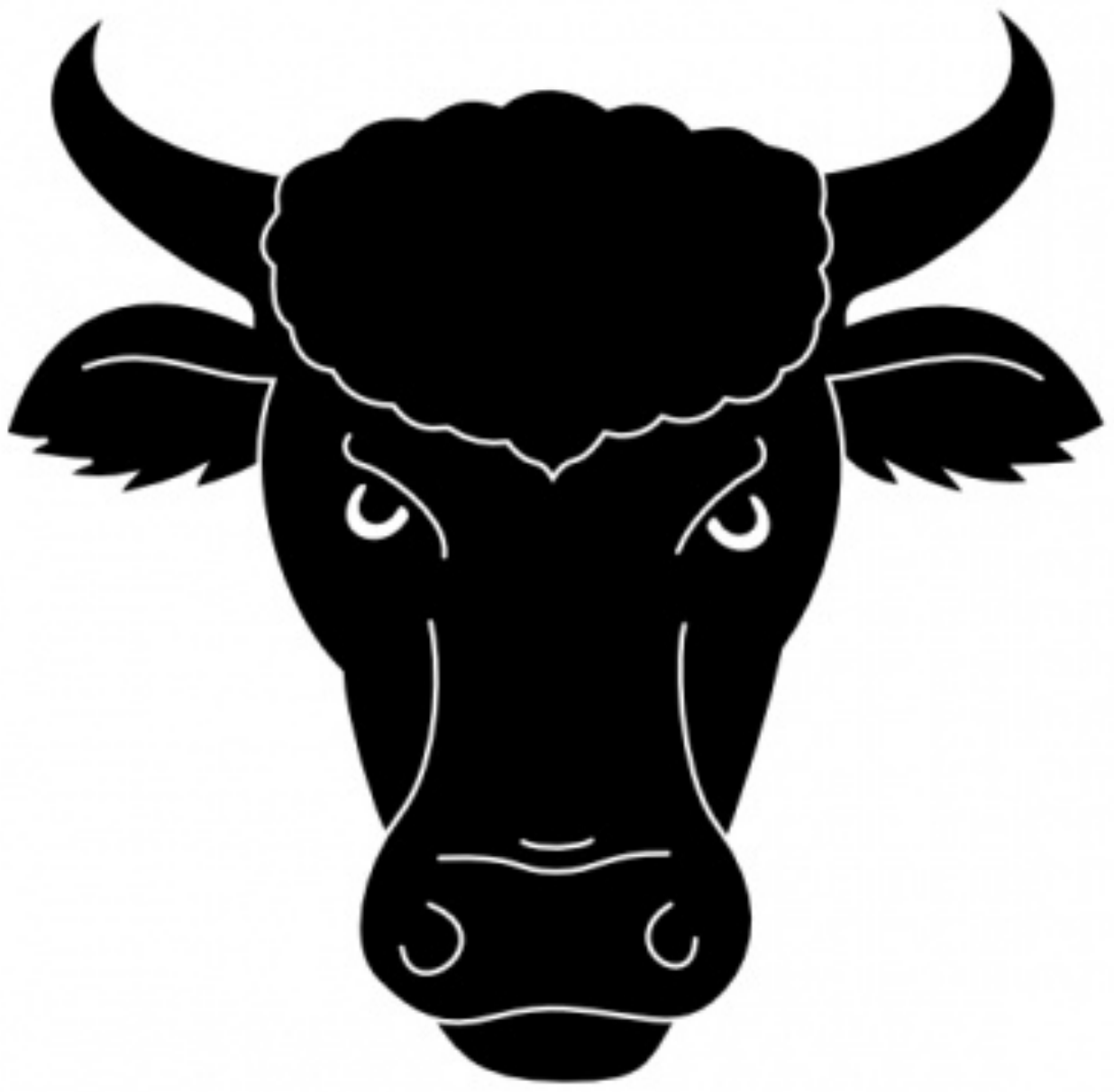}}
\newcommand{\canada}{\includegraphics[height=3.725mm, trim = 0mm 0mm 0mm 0mm, clip]{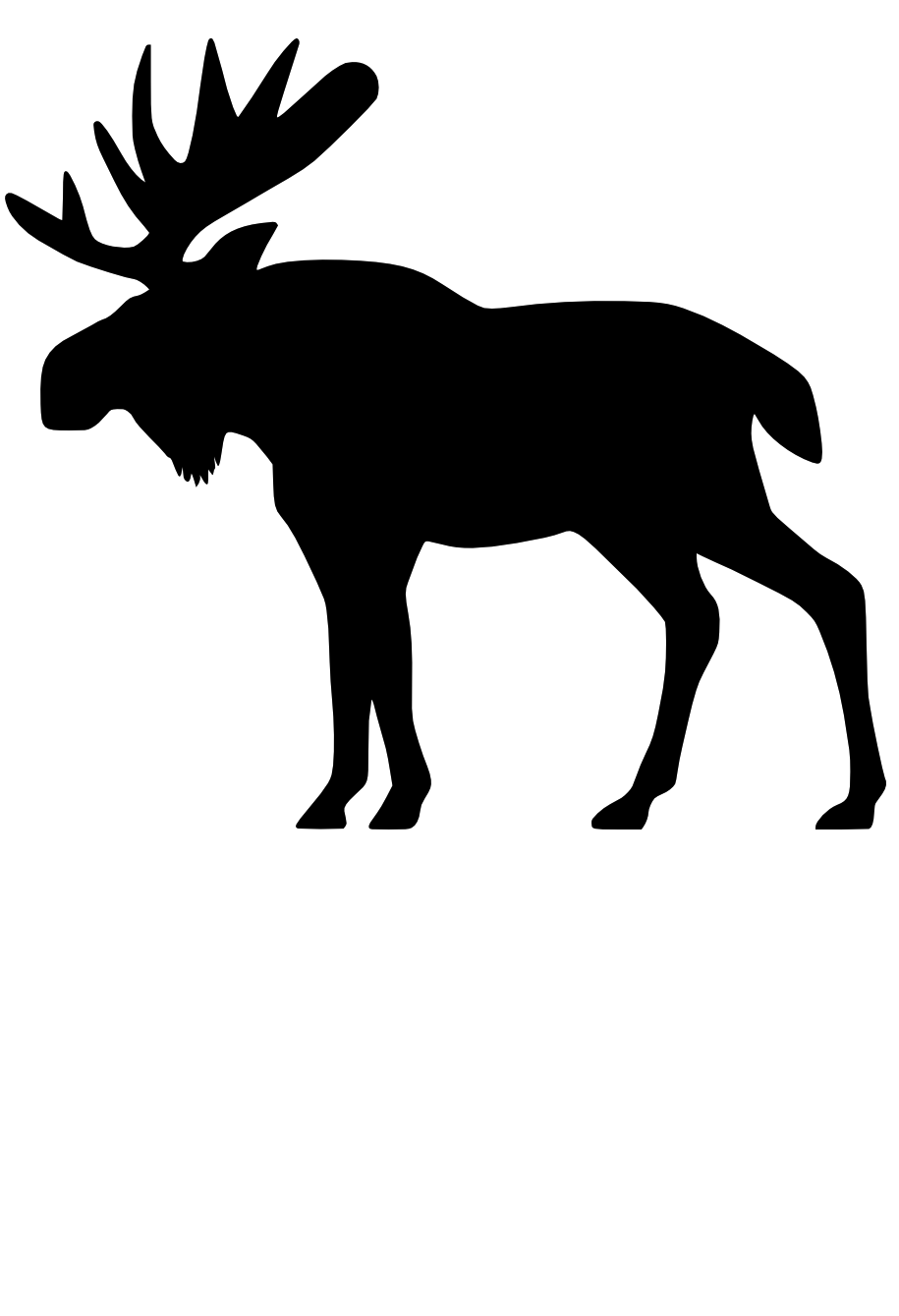}}
 
\begin{document}

\title{\vspace{-1cm}
	\begin{flushright}\end{flushright}
	\bf{A new handle on three-point coefficients:
		\\ \vspace{6pt}
		OPE asymptotics from genus two modular invariance
		}
	\vspace{12pt}}

\author{
	John Cardy\cali\ox,~
	Alexander Maloney\canada  
~and~Henry Maxfield\canada
\\
\\
 \cali Department of Physics, University of California
	\\ 
	Berkeley, CA 94720, USA
	\\
	\\
   \ox All Souls College, Oxford OX1 4AL, UK
	\\
	\\
	\canada Physics Department, McGill University
	\\
	Montr\'eal, QC H3A 2T8, Canada
	\\
	\\
}
\maketitle
%


\vspace{-1em}

\abstract{
We derive an asymptotic formula for operator product expansion coefficients of heavy operators in two dimensional conformal field theory.
This follows from modular invariance of the genus two partition function, and generalises the asymptotic formula for the density of states from torus modular invariance.  The resulting formula is universal, depending only on the central charge, but involves the asymptotic behaviour of genus two conformal blocks.  We use monodromy techniques to compute the asymptotics of the relevant blocks at large central charge to determine the behaviour explicitly.
}


\clearpage

\tableofcontents

\section{Introduction}
Two dimensional conformal field theories are highly constrained by symmetry.  
Remarkably, unitarity and conformal symmetry impose strong non-perturbative constraints on their spectrum and interactions.
 In addition to the infinite dimensional algebra of local conformal symmetries \cite{Belavin:1984vu}, which constrains correlation functions of local operators, modular invariance of CFT partition functions leads to new and a-priori distinct constraints.
For example, modular invariance of the torus partition function can be used to understand the asymptotic density of states \cite{Cardy:1986ie}, to derive bounds on the spectrum \cite{Hellerman:2009bu, Friedan:2013cba, Collier:2016cls} and to determine the asymptotics of certain OPE coefficients \cite{Kraus:2016nwo}.  This implementation of the conformal bootstrap using modular invariance leads to qualitatively different results from approaches based on crossing symmetry of local correlation functions (as in \cite{Rychkov:2016iqz, Simmons-Duffin:2016gjk} and references therein). 

The basic dynamical data of a two dimensional conformal theory is a list of scaling dimensions $\Delta_i$ and spins $J_i$ for the primary operators of the theory, along with a list of three point coefficients $C_{ijk}$ which appear in the operator product expansion of primary operators.  This data completely determines all correlation functions of the theory, as well as the partition function on any surface $\Sigma$.  The CFT partition function on a surface $\Sigma$ will be a function of the conformal structure moduli of $\Sigma$, and must transform appropriately under modular transformations.  These modular symmetries are ``large" conformal transformations of $\Sigma$ -- transformations which are not continuously connected to the identity -- so lead to different constraints from the Virasoro symmetries generated by infinitesimal conformal transformations.  We will work in Euclidean signature, where these modular symmetries are easier to understand.
Our goal is to understand the constraints of modular invariance of the CFT on higher genus Riemann surfaces.  We will focus on genus two, although many of our results can be generalized to higher genus.  

Our motivating example will be the derivation of the asymptotic density of states from torus modular invariance \cite{Cardy:1986ie}. 
We will now give a very schematic review of this result (various details will be made precise later on) in order to outline our general strategy.
Let us consider a conformal field theory on a Euclidean torus.
We will think of this torus as two spheres glued together by a pair of long cylinders, as in Figure 1(a).  
The partition function can be computed by summing over all possible states which propagate along these two cylinders.
If we take each cylinder to have length $\beta/2$, and radius one, then the partition function can be written schematically as
\begin{equation}\label{bilbo}
Z_{g=1}(\beta) \simeq \sum_{i,j} \left(g_{ij}\right){}^2 e^{-\beta (E_i +E_j)/2}~.
\end{equation}  
The sum is over all states $i$ and $j$ propagating along the cylinder, and $E_i=\Delta_i-c/12$ is the energy of the state $i$ on a circle, including the (negative) Casimir energy proportional to the central charge. The exponential suppression factors in \cref{bilbo} are just the amplitudes for these states to propagate a distance $\beta/2$.  Here $g_{ij}$ is the two point function of the states $i$ and $j$ on the sphere: two factors of $g_{ij}$ appear in the partition function, one for each sphere.  Since we can work in a basis where these two point functions are diagonal ($g_{ij} = \delta_{ij}$), we see that $Z(\beta) = \sum_i e^{-\beta E_i}$ is the usual canonical ensemble partition function of the theory at temperature $\beta^{-1}$.  

 \begin{figure}
 \centering
	\begin{subfigure}[b]{.4\textwidth}\centering
	 \includegraphics[width = 0.8\textwidth]{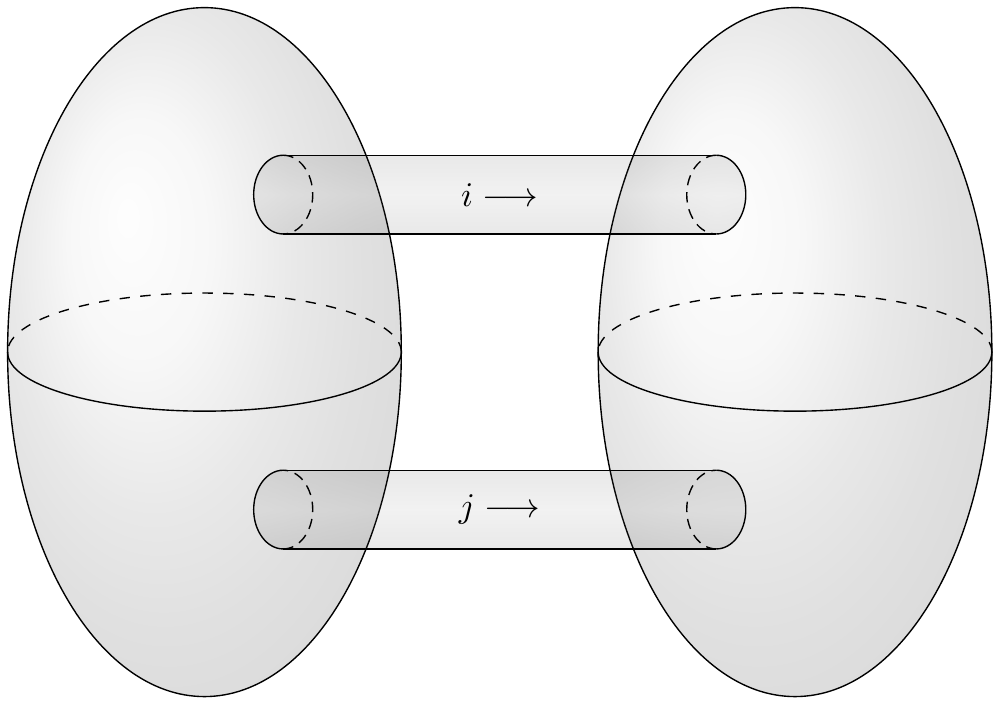}
	 \caption{Genus one}
	\end{subfigure}
	\begin{subfigure}[b]{.4\textwidth}\centering
	 \includegraphics[width = 0.8\textwidth]{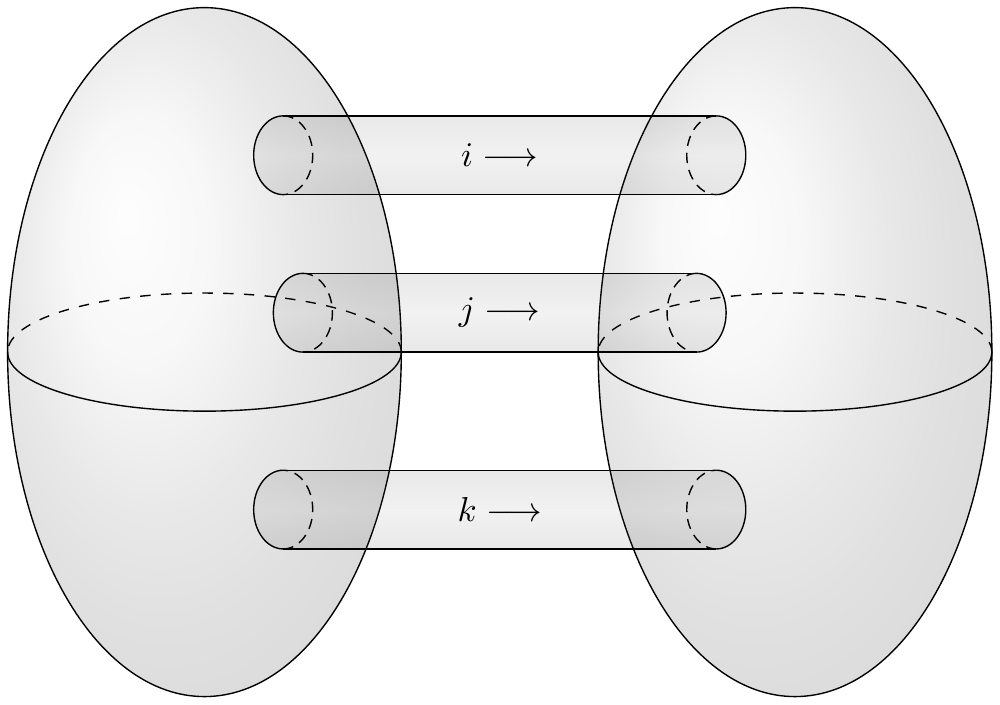}
	\caption{Genus two}
	\end{subfigure}
\caption{\label{fig1}Surfaces constructed by gluing together two spheres with cylinders.  The partition function is then a sum over states labelled $i,j,k$ propagating along the cylinders, times the square of a two- or three-point function on the sphere.}
 \end{figure}
 
Modular invariance is the statement that the conformal structure of this torus is invariant under $\beta \to 1/\beta$, i.e. $Z(\beta) = Z(1/\beta)$.  This relates the low energy behaviour of the spectrum to the high energy behaviour of the spectrum.  For example, it means that the high energy behaviour of the spectrum is determined completely by the energy of the ground state $E_0 = -c/12$.  
In fact, one can view $\beta$ as a complex parameter (the conformal structure modulus of the torus), and so obtain the asymptotic density of states as a function of the left- and  right-moving dimensions
$h=({\Delta+J})/2$ and ${\bar h} = (\Delta-J)/2$ separately.
The result is the asymptotic formula \cite{Cardy:1986ie}
\begin{equation}\label{cardy}
\rho(h, {\bar h}) \approx \exp\left[2\pi \sqrt{{c	\over 6}~ h}+2\pi \sqrt{{c	\over 6}~ {\bar h}}\right]
\end{equation} 
for the density of states at high energy.
   
To generalize this to higher genus, we will think of the genus two partition function as two spheres glued together by three cylinders, as in Fig 1(b).  This leads to a similar schematic expression for the genus two partition function 
\begin{equation}\label{frodo}
Z_{g=2}(\beta) \simeq \sum_{i,j,k} \left(C_{ijk}\right)^2 e^{-\beta (E_i +E_j+E_k)}
\end{equation} 
where now we have called the length of the cylinders $\beta$.  The main difference is that now the coefficient appearing in this sum is the square of the three point coefficient $C_{ijk} \simeq \langle O_i O_j O_k\rangle$ for operators on the sphere.  Since $C_{ijk}$ is not diagonal, we must include all states $ijk$ in the sum.   
The crucial (and not entirely obvious) statement is that modular invariance at higher genus works just as at genus one: it takes $\beta \to 1/\beta$ in \rref{frodo}.  As with the torus partition function, this means that the high energy behaviour of of the theory is determined by the low energy behaviour.  In particular, the asymptotic behaviour of the $\left(C_{ijk}\right)^2$ is determined by the three point functions of the lightest operator (i.e. the identity), which has $C_{111}=1$. This will lead to an asymptotic formula for the three point coefficients. 

There are many subtleties which have been ignored above. 
We have been cavalier about exactly how one constructs the partition function by gluing surfaces together, and ignored several subtleties about how modular invariance acts on the partition function at genus two and how the coefficients $C_{ijk}$ in \cref{frodo} are related to the usual OPE coefficients.  
Most importantly, we will also need to understand the behaviour of the conformal blocks for the genus two surface, which (unlike the genus one case) can not be computed exactly.  Once the dust settles, we will arrive at a result for the average value of the squared three point function coefficient of heavy primary operators, valid when $\sqrt{h_1}+\sqrt{h_2}>\sqrt{h_3}$ (and permutations)\footnote{We will write many asymptotic formulae, so here briefly comment on our notational conventions. We reserve $A\sim B$ for the precise sense that $A/B\to 1$ in the appropriate limit. We will instead write $A\approx B$ to mean something weaker, that $A$ scales like $B$ up to corrections that are less important than what is written (e.g., neglecting order one factors).}:
\begin{equation}\label{saruman}
\overline{\left(C_{h_1 h_2 h_3}\right)^2} \approx  \mathcal{F}_0^{-\left({h_1+h_2+h_3}\right)}\bar{\mathcal{F}}_0^{-\left(\bar{h}_1 + \bar{h}_2 + \bar{h}_3\right)} \exp\left[-\pi\sum_k\left(\sqrt{\frac{c}{6} h_k}+\sqrt{\frac{c}{6} \bar{h}_k}\right)\right] ~.
\end{equation}
The average here is over all primary operators $\opO_i, \opO_j, \opO_k$ of fixed (large) dimensions $h_1,h_2,h_3$, and $\mathcal{F}_0$ is a constant which depends only on the ratios of the dimensions in the limit where the $h_k$ are large.  This constant can be evaluated either numerically, or perturbatively around the point where the dimensions are equal, where we have
\begin{equation}
\blockF_0 =\frac{16}{27}\left( 1+ \log 3 \frac{(h_1-h_2)^2+(h_2-h_3)^2+(h_3-h_1)^2}{4h^2} + \cdots  \right)
\end{equation} 
where $h$ is any one of the dimensions (it does not matter which at this order in the perturbation), or at the edge of the regime of validity $\sqrt{h_1}+\sqrt{h_2}=\sqrt{h_3}$, when $\blockF_0=1$ so there is no exponential piece.
Thus, at fixed total dimension $h_1+h_2+h_3$, the three point coefficient is maximised when the dimensions are equal, where we have
\begin{equation}
\overline{\left(C_{h h h}\right)^2} \approx \left(\frac{27}{16}\right)^{3(h+{\bar h})}\exp\left[-3\pi\left(\sqrt{\frac{c}{6} h}+\sqrt{\frac{c}{6} \bar{h}}\right)\right]~.
\end{equation}
The exponential factor here comes from our choice of conventions; it would be absent, for example, if we chose to place the operators at the vertices of an equilateral triangle with sides of length $27/16$, rather than at $0$, $1$ and $\infty$.  
These expressions are derived for large central charge, where the asymptotic behaviour of the blocks is relatively easy to understand.  The generalisation to finite central charge involves more detailed information about the conformal blocks.

We note that general asymptotic formulas for OPE coefficients can be obtained in other contexts as well.  For example, crossing symmetry of four point functions determines to the asymptotic behaviour of the OPE coefficients when one operator is taken to be heavy and the other two operators are held fixed \cite{Pappadopulo:2012jk}.
Similarly, modular covariance of torus one-point functions determines the asymptotics when one operator is held fixed and the other two operators are taken to be heavy and equal \cite{Kraus:2016nwo}.
It is only genus two modular invariance that constrains the limit where all of the operators are taken to be heavy.

There are several natural generalizations of our approach.  First, although we have presented results at genus two, this approach works at all genus.  The result is an asymptotic formula similar to \cref{saruman} for $n$-point functions of heavy operators on the sphere.  These $n$-point functions are in principle already determined by three point coefficients using the usual OPE, so it would be interesting to see if it is possible to extract useful constraints from this result.  Second, although we have focused on asymptotic properties of three point coefficients, it is also possible to implement a modular bootstrap program by studying the expansion of these partition functions around the modular invariant point.  This is a generalization of the modular bootstrap program on the torus, but requires knowledge of the higher genus conformal blocks.  We are no longer in the asymptotic regime, so one must compute the blocks explicitly order by order in perturbation theory.  This will be discussed in \cite{ccy}.  Finally, one could consider special classes of conformal field theories where this higher genus bootstrap program can be carried out exactly.  For example, in chiral CFTs the constraints of modular invariance are much more powerful, since the space of modular invariant partition functions is finite dimensional.  Thus even knowing the blocks perturbatively is sufficient to obtain exact results on OPE coefficients.  This will be discussed in \cite{kmz}.

We should also comment on the holographic interpretation of these results in terms of AdS$_3$ gravity.  A heavy CFT state (i.e. a state with $\Delta \gg c/12$) is dual to a black hole microstate in AdS.  Indeed, the density of states \cref{cardy} matches precisely the Bekenstein-Hawking entropy of the corresponding black hole in AdS$_3$ \cite{Strominger:1997eq}.  This is because the CFT partition function on the torus is, at high temperature, the Euclidean action of the AdS$_3$ black hole.
An OPE coefficient involving heavy operators should therefore be given a bulk interpretation as correlation function of black hole states.  For example, a light-heavy-heavy three point function can be interpreted as the one-point function of a light operator in the black hole background dual to the heavy state.  This can be computed explicitly in the bulk, and compared to the corresponding asymptotic formula from modular invariance \cite{Kraus:2016nwo}. \footnote{Similar considerations involving four point functions were described in \cite{Fitzpatrick:2014vua}.}
 The heavy-heavy-heavy three point functions considered in this paper can, similarly, be interpreted as a coupling between three black hole states.  
The difference is that this can no longer be interpreted in terms of a one-point function in a fixed black hole background.
Instead, one should interpret the genus two partition function as the Euclidean action of the mutli-black hole solution of AdS$_3$ \cite{Brill:1995jv}.  This is a bulk solution with three asymptotic boundaries, and the multi-black hole state is interpreted as an entangled state in the Hilbert space of three copies of the CFT on a circle.  The three point coefficient can then interpreted in terms of the wave-function of this three-black hole state, as in \cite{Krasnov:2000zq, Skenderis:2009ju, Balasubramanian:2014hda}.  It would be interesting to explore this further.

Our plan is as follows.  In the next section we will discuss the computation of higher genus CFT partition functions in terms of twist operators in an orbifold CFT.  This is somewhat simpler than a construction of the partition function by gluing together surfaces, and has the advantage that modular invariance can be interpreted as crossing symmetry of the twist operator correlation functions.  In section 3 we will show how this leads to asymptotic constraints on OPE coefficients, and rederive the asymptotic formula for the density of states \cref{cardy} using the twist operator language.  We will discover that knowledge of the conformal blocks is absolutely crucial. In section 4 we will apply Zamolodchikov's monodromy technique to study the asymptotic behaviour of the relevant conformal blocks.  This will reproduce known results at genus one, and lead to new results at higher genus.  This will lead to our desired asymptotic formula for OPE coefficients in CFTs with large central charge.

\paragraph{Note:} The submission of this paper is coordinated with those of \cite{ccy} and \cite{kmz}, which explore related aspects of the conformal bootstrap at genus two.

\section{CFT partition functions and twist operator OPE}

	We consider a two dimensional CFT $\CFT$ with central charge $c$.  
	We wish to compute the partition function of the conformal field theory on a Riemann surface $\Sigma$ of genus $g$.  One way to do so is to construct the surface explicitly by a cutting and sewing procedure, as in \cite{Mason:2006dk} (for more details on this approach, see \cite{Gaberdiel:2010jf, Headrick:2015gba, Belin:2017nze}).  We will instead find it more convenient to write the partition function as the expectation value of twist operators in an orbifold CFT, following \cite{ Dixon:1986qv, Knizhnik:1987xp, Lunin:2000yv, Witten:2007kt, Calabrese:2009qy, Calabrese:2009ez}.   One advantage of this approach is that the modular symmetries of the surface $\Sigma$ can be represented as crossing symmetries of the twist operators. 
	
	\subsection{CFT partition function as twist operator correlation function}

A simple way to construct a Riemann surface is to define it as an algebraic curve, as the set of solutions to an equation of the form
\begin{equation}\label{branched}
y^n = \prod_{k=1}^{N} \frac{z-u_k}{z-v_k}\;.	
\end{equation}
We will denote the corresponding surface $\Sigma_{n,N}(u_i, v_i)$, or just $\Sigma_{n,N}$, representing $\Sigma_{n,N}$ as a $n$-sheeted cover of the Riemann sphere $\CC^*$.  The points on $\CC^*$ are parameterized by the coordinate $z$, and the $n$ sheets are labelled by a choice of the root $y$ which solves \rref{branched}.  
This covering map has $2N$ branch points at $(u_k,v_k)$; the monodromy of $z$ around one of these points will shift $y\to e^{\pm 2\pi i/n} y$ and move us from one sheet to another.  The genus of the surface $\Sigma_{n,N}$ is given by the Riemann-Hurwitz formula:
\begin{equation}
g= (N-1)(n-1)\; .
\end{equation}
The locations of the branch points $(u_k,v_k)$ become the moduli of the Riemann surface $\Sigma_{n,N}$.  However, it is in general not the case that the  
$(u_k,z_k)$ will map out the full moduli space of genus $g$ Riemann surfaces.  This is because
all curves constructed in this manner will have $\ZZ_n$ symmetry (usually referred to as replica symmetry) generated by $y\to e^{2\pi i/n} y$.  The $(u_k,v_k)$ can be thought of as coordinates on the moduli space of genus $g$ surfaces with $\ZZ_n$ replica symmetry.

Let us consider first the case where $N=2$.  Then we can use $SL(2,\CC)$ symmetry to map the branch points $(u_1,v_1,u_2,v_2)$ to $(0,x,1,\infty)$, where the cross-ratio $x$ is 
\begin{equation}\label{crossratio}
x={(u_1-v_1)(u_2-v_2)\over (u_1-u_2)(v_1-v_2)}
\end{equation}
and denote the surface as $\Sigma_{n,2}(x)$.  The equation for $\Sigma_{n,2}(x)$ is 
\begin{equation}\label{4branched}
y^n= {z (z-1)\over z-x}~.
\end{equation}
When $n=2$ this surface is a torus, and the parameter $x$ is related to the usual conformal structure parameter of the torus.  When $n=3$ this is a genus two Riemann surface, and the cross-ratio $x$ parameterises a one-dimensional subspace of the three (complex) moduli of genus two surfaces.

This description allows us to represent the partition function of $\CFT$ on $\Sigma_{n,N}$ as a correlation function in the orbifold theory $\CFT^{\otimes n} / \ZZ_n$.  In this description we consider $n$ copies of our original CFT $\CFT$, one living on each sheet. 
If we denote by $\opO_i$ a local operator in the original CFT, untwisted operators in the product theory will be $\ZZ_n$ invariant linear combinations of operators like $\opO_{i_1}^{(1)}\otimes\dots\otimes\opO_{i_n}^{(n)} $, where the superscript indicates which sheet the operator lives on.  
As our surface is $\ZZ_n$ invariant, we will consider only those operators which invariant under the cyclic $\ZZ_n$ permutations of the sheets. 
The orbifold theory will also have twisted sectors. 
In particular, the twist and anti-twist operators $\sigma_n$ and $\bar{\sigma}_n$ have the property that monodromy around $\sigma_n$ moves a local operator from the sheet $i$ to the sheet $i+1$ or $i-1$ respectively.  These twist operators have conformal dimension 
\begin{equation}
\Delta_{\sigma_n}= {c\over 6}\left(n-{1\over n}\right)~.
\end{equation} 
When $n=2$, the twist operator and anti-twist operators are the same ($\sigma_2 = {\bar \sigma}_2$).

The partition function of $\CFT$ on $\Sigma_{n,N}$ can then be written in terms of the correlation function of these twist operators.  
In doing so, there is one more important subtlety, which is related to the conformal anomaly.  The conformal anomaly implies that the partition function $Z(\Sigma_{n,N})$ will depend not just on the conformal structure of $\Sigma$, but also on the choice of metric for $\Sigma$ within a conformal class (i.e. a choice of conformal frame).  When we write the surface $\Sigma_{n,N}$ as a branched cover, we are naturally computing 
using the flat metric $ds^2=dz\,d{\bar z}$ on each of the $n$ sheets, with conical singularities of opening angle $\frac{2\pi}{n}$ at the insertion points of the twist operators $\sigma_n$ (regulated by some means, which is irrelevant once the twist operators are normalised in a canonical way).  The partition function in a different frame -- such as one where the metric is smooth -- will then be related to the twist operator correlation function by a conformal anomaly factor. 
This factor depends only on the central charge and the choice of conformal frame, and
can be computed in terms of a Liouville action.  The result is that the partition function on the surface \cref{4branched} will take the form
\begin{equation}\label{Zsigma}
Z(\Sigma_{n,N}) = e^{c\, S_\text{anomaly}} \; \langle \sigma_n(u_1) {\bar \sigma}_n(v_1) \cdots \sigma_n(u_N) {\bar \sigma}_n (v_N) \rangle~.
\end{equation} 
Here the factor $S_\text{anomaly}$ depends on the choice of conformal frame, but not on any details of the conformal field theory.
We will not actually need this function in order to extract the constraints of modular invariance, since we can instead work directly with the twist operator correlation function.

\subsubsection*{The case $g=1$}

Consider first the case $N=2$ and $n=2$, so that $\Sigma_{2,2}$ is a torus.  
Then the cross ratio $x$ in \cref{4branched} is related to the usual torus conformal structure parameter $\tau$ by
\begin{equation}\label{tauis}
x=\left(\frac{\vartheta_2(\tau)}{\vartheta_3(\tau)}\right)^4,\qquad \tau = i \frac{{}_2F_1\left(\frac{1}{2},\frac{1}{2};1;1-x\right)}{{}_2F_1\left(\frac{1}{2},\frac{1}{2};1;x\right)}.
\end{equation}
As $x$ varies over the complex plane, including changing branches of the hypergeometric functions, the $\tau$ parameter maps out the entire upper half plane, and the entire genus one moduli space is covered\footnote{In particular, since every torus has a $\ZZ_2$ symmetry it is possible to write it as a two-fold branched cover.}. In the limit $x\to 0$, $\tau\to i \infty$, and as $x\to 1$, $\tau \to 0$.  More precisely, if we denote by $q=e^{2\pi i \tau}$ the usual elliptic nome, then we have
\begin{equation}
q \sim \left({x\over 16} \right)^2\quad\text{as }x\to 0,\quad\text{and}\quad q \sim \exp\left( 2\pi^2 \over \log (1-x)\right)\quad\text{as }x\to 1.
\end{equation}

The torus partition function is equal to the finite temperature partition function of the CFT on a circle:
\begin{equation}
Z(\tau, {\bar \tau}) = \Tr q^{L_0-{c\over 24}} {\bar q} ^{{\bar L}_0-{c\over 24}}
\end{equation} 
This partition function is computed with the usual (non-singular) flat metric on the torus, so to compare to the twist operator computation it is necessary to compute the conformal anomaly factor for this change of conformal frame, which gives \cite{Lunin:2000yv}
\begin{equation} \label{ZTorus}
\langle\sigma_2(0)\sigma_2(x)\sigma_2(1)\sigma_2(\infty)\rangle= \left|2^8x(1-x)\right|^{-c/12}\,Z(\tau,\bar{\tau})\;.
\end{equation}

\subsubsection*{The case $g=2$}

The other example we are concerned with is the genus two surface.  A simple family of genus two surfaces can be constructed as the $N=2$ curve \cref{4branched} with $n=3$.  As the cross-ratio $x$ varies over the complex plane, we map out a one complex-dimensional slice of the moduli space of genus $g=2$ Riemann surfaces, restricted to surfaces with $\ZZ_3$ symmetry.

One useful way to characterize the conformal structure of these surfaces is through their period matrices, which is the natural generalization of the $\tau$ parameter to higher genus. 
This was computed in \cite{Calabrese:2009ez} (in a canonical basis of cycles):
\begin{equation}\label{genustwo}
\Omega = {1\over \sqrt{3}}\begin{pmatrix}2 & -1 \\ -1 & 2\end{pmatrix} \tau_3, \quad \text{where}\quad  \tau_3 = i \frac{{}_2F_1\left(\frac{1}{3},\frac{2}{3};1;1-x\right)}{{}_2F_1\left(\frac{1}{3},\frac{2}{3};1;x\right)}.
\end{equation}
The $\tau_3$ parameter takes values in the upper half plane, as with the usual torus modular parameter, so that the period matrix $\Omega$ has positive imaginary part. Just as in the torus case, $\tau_3$ is pure imaginary for real $x$ between $0$ and $1$, and $\tau_3\to i \infty$ as $x\to 0$, and $\tau_3 \to 0$ as $x\to 1$.  

A genus $g=2$ Riemann surface can also be represented as a 2-fold cover of the plane branched over six points, i.e.~as \cref{branched} with $N=3$ and $n=2$.  In this presentation, the partition function $Z(\Sigma_{3,2})$ is the 6 point function of twist two operators: 
\begin{equation}
\left\langle \prod_{k=1}^3 {\sigma}_2(u_k){\bar \sigma_2}(v_k)\right\rangle
\end{equation} 
In fact, since every genus two surface has a (unique) $\ZZ_2$ symmetry with six fixed points (a `hyperelliptic involution'), it is possible to represent every genus two surface as a two-fold cover of this form (as a `hyperelliptic curve').
Thus as the $(u_k, v_k)$ are varied we map out the full moduli space of genus two curves.  For example, three of the six complex parameters $(u_k,v_k)$ can be set to $(0,1,\infty)$ using $SL(2,\CC)$ invariance, and the three remaining cross-ratios can be regarded as the conformal structure moduli of $\Sigma_{3,2}$. 

These two different constructions of a genus two surface can be compared if we arrange the six twist two operators in a configuration which has an additional $\ZZ_3$ symmetry.  For example, we can place the branch points on the unit circle at
\begin{equation}\label{6pt}
u_k = \exp\left( {2 \pi i k + i \theta  \over 3}\right) \qquad
v_k = \exp\left( {2 \pi i k - i \theta  \over 3}\right)\; .
\end{equation} 
Then the Riemann surface $\Sigma_{2,3}(u_k, v_k)$ has the same conformal structure as the surface $\Sigma_{3,2}(x)$ constructed using twist three operators when the cross ratio $x$ is related to $\theta$ by
\begin{equation}
x= \cos^2 \frac{\theta}{2}.
\end{equation}
The partition functions are not quite equal, however, since they are evaluated in different conformal frames.  One has to to compute the conformal anomaly factor for the map from the flat metric with six conical defects on $\Sigma_{2,3}$ to the flat metric with four conical defects on $\Sigma_{3,2}$.  The conformal anomaly factor is not needed for our results, so we will not give the details of the calculation here, but for completeness give the result: 
\begin{equation}
\left\langle \prod_{k=1}^3 {\sigma}_2(u_k){\bar \sigma}_2(v_k) \right\rangle = \left(\frac{3}{4}\right)^{3 c/4}(x(1-x))^{5c/72} \langle \sigma_3(0)\bar{\sigma}_3(x) \sigma_3(1) \bar{\sigma}_3(\infty)\rangle
\end{equation}

\subsection{Crossing symmetry as modular symmetry}

The CFT partition function on a Riemann surface $\Sigma$ must transform in an appropriate way under modular transformations.  At genus one, this constrains the spectrum of the theory, but at higher genus this constrains the OPE coefficients as well.  
This modular invariance has a simple interpretation in the language of orbifold CFT: it is the crossing symmetry of twist operator correlation functions.

For example, in a $\ZZ_n$ orbifold theory the twist operator four point function 
\begin{equation}
\langle \sigma_n(u_1) {\bar \sigma}_n(v_1) \sigma_n(u_2) {\bar \sigma}_n (v_2) \rangle\;,
\end{equation}
is invariant under the permutation $u_1\leftrightarrow u_2$, which from \cref{crossratio} takes the cross-ratio $x\to 1-x$.
To see the relation with modular invariance, consider first the case $n=2$.  
We see from \cref{tauis} that $x\mapsto 1-x$ acts on the torus modular parameter $\tau$ parameter as the usual modular $S$ transformation
\begin{equation}
S: \tau\mapsto-1/\tau
\end{equation} 
associated with the element 
\begin{equation}\label{fingolfin}
s=\begin{pmatrix}0&1\\-1&0\end{pmatrix}\in SL(2,\ZZ)~.
\end{equation}

This immediately generalizes to higher genus.  For example, if we represent the genus two surface as a four point function of twist-three operators, the crossing transformation $x\mapsto1-x$ acts on the $\tau_3$ modular parameter as
\begin{equation} 
S:\tau_3\mapsto-{1/ \tau_3}
\end{equation} 
We see that the conformal structure modulus $\tau_3$ is playing exactly the same role for genus two surfaces as the $\tau$ parameter does for torus.
If we instead describe the genus two surface as a six-point function of twist two operators at the points \cref{6pt}, then the crossing transformation is $\theta \mapsto \pi-\theta$.

This transformation can be represented directly as a genus two modular transformation as follows. The genus two modular group $Sp(4,\ZZ)$ 
is the set of linear transformations on the cohomology group $H^1(\Sigma,\ZZ)$ which leaves invariant the symplectic pairing 
 $\int \alpha \wedge \beta$ between cocycles. 
Under a modular transformation 
 \begin{equation}
 M =\begin{pmatrix}D&C \\ B&A\end{pmatrix} \in Sp(2g,\ZZ)
 \end{equation} 
the period matrix  $\Omega$ transforms as
 \begin{equation}
 M: \Omega \to (A\Omega +B) (C \Omega + D)^{-1}~.
 \end{equation} 
From \cref{genustwo}, it is straightforward to check that the modular transformation
\begin{equation}
S = \begin{pmatrix}0&-(s^{-1})^t\\ s&0\end{pmatrix} ,
\end{equation} 
takes $\tau_3\mapsto -1/\tau_3$, where $s$ is as in \cref{fingolfin}.  This is the genus two analog of modular $S$-invariance on the torus.  This is the S-transform that will lead to our desired asymptotic formula for OPE coefficients.

\subsubsection*{Relation to R\'enyi entropies}

This description of higher genus partition functions as correlation function of twist operators is especially useful in computations of R\'enyi entropies in two dimensional CFTs.
For a pair of intervals with endpoints $[u_1,v_1]$, $[u_2,v_2]$, the R\'enyi mutual information
\begin{equation}
I^{(n)}(x) =S^{(n)}([u_1,v_1]) + S^{(n)}([u_2,v_2]) - S^{(n)}([u_1,v_1] \cup [u_2,v_2])
\end{equation} 
is a function only of the cross-ratio $x$.  Here
$S^{(n)}(x) = {1\over 1-n} \log \Tr \rho^n$ is the R\'enyi entropy of the reduced density matrix $\rho$ for the region $[u_1,v_1]\cup [u_2,v_2]$ in the vacuum state.  This can be computed using the replica trick, where it is related to the partition function of the CFT on a $n$-fold cover of the sphere branched at $(u_k, z_k)$ \cite{Calabrese:2009ez}:
\begin{equation}
I^{(n)}(x) = {1\over n-1} \log\frac{
\langle \sigma_n(0) \bar{\sigma}_n(x) \sigma_n(1) \bar{\sigma}_n (\infty) \rangle}
{\langle \sigma_n(0) \bar{\sigma}_n(x)\rangle\langle\sigma_n(1) \bar{\sigma}_n (\infty)\rangle}
\end{equation}
In any pure state, unitarity implies that the R\'enyi entropy of a region is equal to that of its complement.  The complement of the pair of intervals $[u_1,v_1]\cup [u_2,v_2]$ is a pair of intervals with cross-ratio $1-x$.
So the crossing symmetry (i.e. modular invariance) we are studying is a simple consequence of unitarity.  

\section{An asymptotic formula for OPE coefficients}

Our goal is to use modular symmetry to extract the asymptotic behaviour of the 3 point coefficients.  We will first outline the general strategy, using the language of the twist operator OPE, and see that it is necessary to understand higher genus conformal blocks in order to get the correct asymptotics. 
We will work out the genus one case explicitly, where the conformal blocks are easy to compute, and use this approach to rederive the usual formula for the asymptotic density of states \cite{Cardy:1986ie}.
We will then present the result at genus two, using the results for the conformal blocks that will be derived in \cref{blocks}.

\subsection{The general strategy}

We start by considering the OPE expansion of the twist operator four point function:
\begin{equation}\label{SSexpansion}
\langle 
\sigma_n(0)\bar{\sigma}_n(x) \sigma_n(1)\bar{\sigma}_n(\infty)
\rangle
= (x {\bar x})^{-\Delta_{\sigma_n}}
\sum_m 
\left(C_{\sigma_n {\bar\sigma}_n m}\right)^2~ x^{h_m}{\bar x}^{{\bar h}_m}
\end{equation}
The sum here is over operators in the untwisted sector of $\CFT^n/\ZZ_n$, labelled by an index $m$.  We denote the left- and right-moving dimensions of this state by $h_m$ and $\bar{h}_m$, which are related to the total scaling dimension and spin by $\Delta_m=h_m + \bar{h}_m$, $s=h_m-\bar{h}_m$.
An operator in the untwisted Hilbert space of $\CFT^n / \ZZ_n$ is built out of $n$ operators in the original theory, assembled together as
\begin{equation}\label{ois}
\opO_m(z) = \opO^{(1)}_{i_1}(z)\otimes \dots \otimes \opO^{(n)}_{i_n}(z) + \text{cyclic} \;,
\end{equation}
where $\mathcal{O}^{(a)}_{i_a}$ denotes an operator living on the $a$th sheet.  Thus $m$ is collective index $(i_1,\dots,i_n)$, where each $i_a$ labels an operator in $\CFT$.
The dimension of the operator $\opO_m$ is the sum of the dimension of its constituent operators $\opO^{(a)}_{i_a}$:
\begin{equation}
h_m = \sum_{a} h_{i_a},\quad\bar{h}_m = \sum_{a} \bar{h}_{i_a}
\end{equation}

Our argument will have two essential ingredients.  The first is the observation that crossing symmetry constrains the behaviour of the sum (\ref{SSexpansion}) as $x\to 1$, which in turn constrains the asymptotic behaviour of $C_{\sigma_n {\bar \sigma}_n m}$.  To see this, note that the four point function (\ref{SSexpansion}) has a pole of order $|x|^{-2\Delta_{\sigma_n}}$ as $x\to 0$.  This is the contribution of the identity operator in the sum, with $C_{\sigma_n {\bar \sigma}_n 1} =1$.  Combining this with crossing symmetry, we see that the sum must diverge as $x\to 1$:
\begin{equation} \label{asymptotic}
\sum_m \left(C_{\sigma_n {\bar \sigma}_n m}\right)^2 x^{h_m}{\bar x}^{{\bar h}_m} \sim \left|1-x\right|^{-2\Delta_\sigma},\quad\Delta_\sigma=\frac{c}{12}\left(\frac{n^2-1}{n}\right)
\end{equation}
No single term in the sum produces such a singularity, so the only way to reproduce this pole is from many operators with large $\Delta_m$. 
 
Since the states become very dense at high energy, the sum \cref{asymptotic} can be approximated as an integral over a continuous density of states.  Equation \rref{asymptotic} can then be inverted to extract the asymptotic behaviour of the coefficients $C_{\sigma_n {\bar \sigma}_n m}$.   
For example, we can immediately conclude that the contribution from the states at large $(h_m,\bar{h}_m)$
\begin{equation}\label{gollum}
\sum_{(h_m, \bar{h}_m)\text{ fixed}} \left(C_{\sigma_n {\bar \sigma}_n m}\right)^2
\sim {\left(h_m{\bar h}_m\right)^{\Delta_\sigma+1}}
\end{equation} 
must grow polynomially in $h_m$ and $\bar{h}_m$ when these dimensions are taken to be large. This can be seen from the saddle point approximation to the inverse Laplace transformation of \rref{asymptotic}, or by simply noting that \rref{asymptotic} is a binomial series.  

The second essential ingredient in our argument is that the three point coefficients 
$C_{\sigma_n {\bar \sigma_n} m}$ are given by $n$ point correlation functions of operators in the original CFT $\CFT$:
\begin{equation}
C_{\sigma_n {\bar \sigma_n} m} \propto \langle \mathcal{O}^{(1)}_{i_1}\cdots \mathcal{O}^{(n)}_{i_n}\rangle_{S^2}
\end{equation} 
Thus \rref{gollum} is actually a statement about the asymptotic behaviour of the $n$-point correlation functions.
To see this, compute the OPE coefficient from the three-point correlation function:
\begin{equation}
C_{\sigma_n \bar{\sigma}_n m} = \langle \sigma_n(0)\mathcal{O}_{m}(1)  \bar{\sigma}_n (\infty)  \rangle
\end{equation}   
This is the expectation value of the operator \rref{ois}, evaluated on the $n$-fold cover of the sphere branched at two points, $u=0$ and $v=\infty$, with the operators $\opO_{i_a}^{(a)}$ inserted at the points $z=1$ on each of the $n$ different sheets of this branched cover. 
Now, the $n$-fold cover of a sphere branched a two points is just the sphere itself, since we can unwrap this branched structure by defining a coordinate $w=z^{1/n}$.  Then $w$ will be a conventional, single valued coordinate on the Riemann sphere, and on the $w$-plane the operators $O_{i_a}^{(a)}$ will be inserted at the $n^{th}$ roots of unity on the unit circle.
Thus, when the operators $\mathcal{O}_{i_a}^{(a)}$ are Virasoro primaries, we have 
\begin{equation}\label{csigma}
C_{\sigma_n{\bar \sigma}_n m} 
= n^{-(\Delta_{i_1}+\cdots+\Delta_{i_n})} G_{i_1\dots i_n}
\end{equation}
where we define
\begin{equation}\label{moria}
G_{i_1\dots i_n} =\mathcal{N} \left\langle \prod_{a=1}^n O_{i_a}(e^{2\pi i a/n}) \right\rangle_{S^2}
\end{equation}
as the $n$-point function of primary operators arranged on the unit circle, and $\mathcal{N}$ is an order one factor to fix the normalisation of $\mathcal{O}_m$, that depends only on how many of the $\mathcal{O}_{i_a}^{(a)}$ are identical (not their dimensions). The prefactor in \cref{csigma} comes from the conformal transformation from the $z$ coordinate to the $w$ coordinate.

We are now confronted with an important computational subtlety.  When the operators $\opO_{i_a}^{(a)}$ are not primary, they will transform in a complicated way under the map $z\to w=z^{1/n}$ which turns $C_{\sigma_n{\bar \sigma}_n m}$ into an $n$-point function of operators on the unit circle.  
States in the same conformal family will, in general, mix with one another. 
One can compute $C_{\sigma_n{\bar \sigma}_n m}$ explicitly only for low-lying descendants.  In general, all we know is that if the
$\opO_{i_a}^{(a)}$ are descendants of some set of primary operators, then  
$C_{\sigma_n{\bar \sigma}_n m}$ will be proportional to the OPE coefficient of those primary operators arranged on the unit circle, as in \rref{moria}. 
 
The result is that we can organize the sum over states into a sum over conformal families, and rewrite \rref{asymptotic} as a sum over primary operators:
\begin{equation}\label{sauron}
\sum_{i_1\dots i_n \atop primary} \left(G_{i_1\dots i_n}\right)^2  |\blockF(h_{i_1},\dots,h_{i_n}; x)|^2 
\sim \left|1-x\right|^{-{c\over 6}\left({n^2-1\over n}\right)}
\end{equation} 
Here the sum is over collections $(i_1,\dots,i_n)$ of primary operators in $\CFT$.  The conformal block $\mathcal{F}(h_{i_1},\dots,h_{i_n}; x)$ describes the total contribution of all operators of the form 
\begin{equation}
\left(\mathcal{L}_{-\bm{m_1}}\opO^{(1)}_{i_1}\right)\otimes \dots \otimes \left(\mathcal{L}_{-\bm{m_n}}\opO^{(1)}_{i_1}\right),
\end{equation} 
where the $\opO_{i_a}^{(a)}$ are primary, and $\mathcal{L}_{-\bm{m_a}}$ denotes a general product of Virasoro raising operators.  Since we can factor these into left- and right-moving conformal blocks, we have written the total block in \rref{sauron} as the absolute value squared of a holomorphic block $ \blockF(h_{i_1},\dots,h_{i_n}; x)$.  We have absorbed the factor of $x^{h_m}$ from \rref{asymptotic}, as well as the factor of $n^{h_m}$ from \rref{csigma} into the conformal block, so that the block has a perturbative expansion
\begin{equation}
\blockF(h_{i_1},\dots,h_{i_n}; x) = x^{-2h_\sigma}\left(x\over n^2\right)^{h_{i_1}+\dots+h_{i_n}} \Big(1+ \opO(x)\Big)
\end{equation} 
at small $x$.  The block $\blockF(h_{i_1},\dots,h_{i_n}; x)$ is a purely kinematic object, which depends only on the weights $h_i$, the cross-ratio $x$ and the central charge $c$.

It is important to note that, even though we are studying four point functions of twist operators, $ \blockF(h_{i_1},\dots,h_{i_n}; x)$ is \emph{not} a standard four point conformal block with respect to the Virasoro algebra of the tensor product CFT $\CFT^{\otimes n}/ \ZZ_n$.  
In particular, there are many operators $\mathcal{O}_m$ which are primary with respect to the Virasoro algebra of $\CFT^{\otimes n}/ \ZZ_n$, but yet are not the tensor product of primary operators in $\CFT$.  
If one wants to interpret $\blockF(h_{i_1},\dots,h_{i_n}; x)$ as a conformal block in the orbifold theory, it should be regarded as a conformal block for the extended algebra of the untwisted sector of $\CFT^{\otimes n}/\ZZ_n$, which is a cyclic product of $n$ copies of the Virasoro algebra.

When $n=2$, equation \rref{sauron} will reproduce the usual formula for the asymptotic density of states.  When $n=3$, it will lead to an asymptotic formula for three point coefficients.

\subsection{Genus one: the density of states}\label{CardyDerivation}

As a warmup, let us begin by considering the genus one case, $n=2$.
With conventional normalisations, the primary operator two point function \rref{moria} is
\begin{equation}
G_{ij} = \left\langle \mathcal{O}_i(-1) \mathcal{O}_j(1)\right\rangle = \delta_{ij} 2^{-2\Delta_i}
\end{equation} 
so the sum \rref{sauron} then reduces to
\begin{equation}\label{saurontorus}
\langle \sigma_2(0)\sigma_2(x) \sigma_2(1)\sigma_2(\infty)\rangle=\sum_{\text{primary }i} \left|\blockF_{g=1}(h_i;x)\right|^2
\sim 
|1-x|^{-c/4}
\end{equation} 
where $\mathcal{F}_{g=1}(h;x)\equiv 2^{-4h}\mathcal{F}(h, h;x)$ is the torus conformal block (absorbing the extra factor from $G_{ij}$ so that $\mathcal{F}_{g=1}(h;x)\sim 2^{-8h}x^{2h-2h_\sigma}$ as $x\to 0$), evaluated in the twist operator frame.  
Near $x=1$, this sum is dominated by the heavy part of the spectrum, so we will need these blocks only when $h$ is large.
These blocks will also have a non-trivial dependence on $x$, which we will need to keep in order to reproduce the correct $x\to 1$ behaviour.  

Happily, it is possible to compute the blocks on the torus exactly.
In the frame where the torus is flat, the partition function is simply the sum over states on the circle
\begin{equation}
Z(\tau, {\bar \tau}) = \sum_{\text{primary }i} 
 \chi_{h_i}(\tau) \bar{\chi}_{\bar{h}_i}(\bar{\tau})
\end{equation} 
where $\chi_{h_i}(\tau)$ is a character of the Virasoro algebra.  Comparing with \rref{ZTorus}, we can read off the conformal blocks in the twist operator frame as
\begin{equation}
\mathcal{F}_{g=1} (h_i,x) = \left(2^8 x (1-x)\right)^{-c/24}\chi_{h_i}(\tau),
\end{equation}   
and we will restrict our attention to the case $c>1$, where the characters
\begin{equation}
\chi_{h_i}(\tau) = q^{h_i-c/24}\prod_{n=1}^\infty {1\over 1-q^n} = \frac{q^{h_i-\frac{c-1}{24}}}{\eta(\tau)}
\end{equation} 
depend trivially on $h_i$. This expression is normalised as expected, with $\mathcal{F}_{g=1}\sim 2^{-8h}x^{-\frac{c}{8}+2h}$ at small $x$.

When $c>1$, there are many heavy primary states, and the sum \cref{saurontorus} can be approximated by an integral:
\begin{equation}
\int dh \; d{\bar h}~ \rho(h, {\bar h})~ \left|\blockF_{g=1}(h,x)\right|^2 
\sim |1-x|^{-c/8}
\end{equation} 
Here $h$ and ${\bar h}$ are left and right-moving dimensions, and $\rho(h, {\bar h})$ is the primary operator density of states.
To reproduce the pole at $x\to1$ we need the asymptotic behaviour of $\blockF_{g=1}(h,x)$.  Since
\begin{equation}
\eta(\tau)\sim (1-x)^{1/12} \quad \text{as } x\to1,
\end{equation} 
we find 
\begin{equation}
\blockF_{g=1} (h,x) \approx (1-x)^{-\frac{c}{24}-\frac{1}{12}} \exp\left[h \left({2\pi^2 \over \log (1-x)} \right)\right]\;.
\end{equation} 
We have kept here only those factors which appear at leading order in $h$, or which contribute to the leading divergence as $x\to1$, and used the asymptotic behaviour of $q$ as $x\to 1$.
The result is that the density of states $\rho(h, {\bar h})$ is determined by
\begin{equation}
\int dh d{\bar h}~ \rho(h, {\bar h})~ 
 \exp\left[h \left({2\pi^2 \over \log (1-x)}\right) + {\bar h} \left({2\pi^2 \over \log (1-{\bar x})}\right) \right]
\sim |1-x|^{-\frac{c-1}{12}}\;.
\end{equation} 
We can then invert this  to obtain the usual expression for the asymptotic density of states \cite{Cardy:1986ie},
\begin{align}
\rho(h, {\bar h}) 
\approx& \int d\beta d {\bar \beta} \exp\left[
\beta h + {\bar \beta} {\bar h} 
+ \pi^2\frac{c-1}{6}\left({1\over \beta} + {1\over {\bar \beta}} \right)
\right]
\\
\approx&  
\exp\left[ 2\pi 
\left(\sqrt{{c-1\over 6} h} + \sqrt{{c-1\over 6} {\bar h}}\right)\right].
\end{align}
In the first line we have written the inverse Laplace transform using the variable $ \beta = -{2\pi^2 \over \log (1-x)}$, and in the second we have made the usual saddle point approximation which is valid as $x\to 1$.  We note that the factor of $c-1$ appears here because we are counting primary states; if we counted all states, not just primaries, then this factor of $c-1$ would be replaced by $c$.

\subsection{Genus two: three point coefficients}\label{OPEderivation}

In generalising this to genus two, there are two possible approaches.  The simplest strategy would be to consider \rref{sauron} with $n=3$.  However, we will take a slightly more general approach, and instead consider the genus two surface as a six point function of twist-two operators inserted at 
\begin{equation}\label{locations}
u_k = e^{i(2\pi k+\theta_k)/3},\quad  v_k = e^{i(2\pi k-\theta_k)/3}
\end{equation} 
with $k=1,2,3$.  
Since we now have three independent variables $\theta_k$, we will obtain a more general result where the dimensions of the three point coefficients appearing in the three-point coefficient can be varied separately.

We now evaluate the six point function
\begin{equation}\label{isengard}
\left\langle \prod_{k=1}^3 \sigma_2(u_k) \sigma_2 (v_k) \right\rangle
\end{equation} 
using the operator product expansion for twist-two operators in the $\CFT^{\otimes 2}/\ZZ_2$ orbifold theory.
In particular, we can choose the channel where we take the OPE between the operators at $u_{k+1}$ and $v_{k-1}$ (with $k$ taken modulo 3), which both approach $-e^{2\pi i k/3}$ as we take $\theta_k\to \pi$:
\begin{equation}
	\sigma_2(u_{k+1})\sigma_2(v_{k-1})= \sum_{m} |u_{k+1}-v_{k-1}|^{\Delta_m-c/4} C_{\sigma_2 \sigma_2 m} \opO_m(-e^{2\pi i k/3})
\end{equation}
We have here suppressed a phase for operators $\opO_m$ with spin, which will not be important in what follows.
Inserting this OPE for the three pairs of twist operators, the correlation function \rref{isengard} becomes a sum over terms like
\begin{equation}\label{shelob}
	C_{\sigma_2 \sigma_2 m_1}  C_{\sigma_2 \sigma_2 m_2}  C_{\sigma_2 \sigma_2 m_3} \left\langle \mathcal{O}_{m_1}(-e^{2\pi i/3})\mathcal{O}_{m_2}(-e^{4\pi i/3}) \mathcal{O}_{m_3}(-1) \right\rangle
\end{equation}
where the $m_k$ run over all the operators in the untwisted sector of the orbifold theory. These operators take the form 
\begin{equation}
	\mathcal{O}_{m_k}(-e^{2\pi i k/3}) = \mathcal{O}^{(1)}_{i_k}\left(-e^{2\pi i k/3}\right) \otimes_\text{Sym} \mathcal{O}^{(2)}_{j_k}\left(-e^{2\pi i k/3}\right)
\end{equation}
where the two factors are operators in the seed theory $\mathcal{C}$ on the two sheets.  We can therefore rewrite \rref{shelob} in terms of three point coefficients in $\mathcal{C}$.  In particular, for primary operators we can use 
$ C_{\sigma_2 {\bar \sigma_2} m_k} = 2^{-4\Delta_{i_k} }\delta_{i_kj_k} $, so that \rref{shelob} is proportional to
the square of the three point function:
\begin{equation}
\left\langle
O_{i_1}(1) O_{i_2} (e^{2\pi i/3}) O_{i_3} (e^{-2\pi i /3}) \right \rangle_{S^2}
 = 3^{-(\Delta_{i_1}+\Delta_{i_2} + \Delta_{i_3})/2} C_{i_1 i_2 i_3}~.
\end{equation} 
We have included here the kinematic factor which relates three point functions on the plane to the usual three point coefficients.

As in the torus case, for non-primary operators the operator product coefficient $C_{\sigma_2 \sigma_2 m_k} $ will mix states within a conformal family in a non-trivial way, but the contributions within a conformal family are still determined kinematically.  Thus we can write  
\begin{equation}\label{samwise}
	\left\langle \prod_{k=1}^3 \sigma_2(u_k) {\bar \sigma}_2 (v_k) \right\rangle = \sum_{\substack{i_1,i_2,i_3\\ \text{primary}}} \left(C_{i_1,i_2,i_3}\right)^2 \left|\mathcal{F}_{g=2}({h_1,h_2,h_3};\theta_k)\right|^2
\end{equation}
where $\mathcal{F}_{g=2}({h_1,h_2,h_3};\theta_k)$ is the appropriate genus two conformal block.
Here we have normalised $\mathcal{F}_{g=2}({h_1,h_2,h_3};\theta_k)$ so that its
 $\theta_k\to\pi$ limit is 
\begin{equation}\label{smallTheta}
 \mathcal{F}_{g=2}({h_1,h_2,h_3};\theta_k) \sim  \frac{1}{48^{h_1+h_2+h_3}} \prod_{k=1}^3 \left(\frac{2\pi-\theta_{k+1}-\theta_{k-1}}{3}\right)^{2h_k-c/8}.
 \end{equation}
This normalization is convenient since it absorbs the factors of $2^{-4h_k}$ from the $C_{\sigma_2 \sigma_2 m_k}$ OPE coefficients, as well as the factors of $3^{h_k/2}$ which arise when we transform the three point function $\langle \opO_{i_1}(0) \opO_{i_2}(e^{2\pi i/3}) \opO_{i_3}(e^{-2\pi i/3})\rangle$ into the conventional three point coefficient $C_{i_1 i_2 i_3}$. 

Just as in the torus case, crossing symmetry now determines the asymptotic behaviour of this sum. In particular, it implies that we may take the OPE in a different channel, such as the one where we fuse the operators $u_k$ and $v_k$ fuse. The appearance of the identity operators in this channel leads to a pole as $\theta_k\to 0$, so we have the singularity
\begin{equation}\label{blockSingularity}
\sum_{\substack{i_1,i_2,i_3 \\ \text{primary}}} \left(C_{i_1,i_2,i_3}\right)^2 \left|\mathcal{F}_{g=2}({h_1,h_2,h_3};\theta_k)\right|^2
\sim \left(\prod_{k=1}^3 \frac{2\theta_k}{3} \right)^{-c/4}\quad\text{as}\quad\theta_k\to0.
\end{equation} 
Note that this sum is organized so that light operators will dominate the sum when $\theta_k\to\pi$.  The singular behaviour \rref{blockSingularity}, which comes from the identity operator in the cross channel, will be reproduced by the sum over heavy operators.

In order for this sum to be useful, we will need to understand the behaviour of the conformal blocks. In the next section, we will show that when $c\gg 1$, the relevant asymptotics at large dimension and $\theta_k\to 0$ of these blocks are given by 
\begin{equation}\label{blockAsymptotics}
	\mathcal{F}_{g=2}(h_1,h_2,h_3;\theta_k)\approx \mathcal{F}_0^{h_1+h_2+h_3} \left(\prod_k\theta_k \right)^{-\frac{c}{24}} \exp\left[-\sum_k \frac{\pi^2}{2\log\frac{1}{\theta_k}}\left(\sqrt{h_{k+1}}+\sqrt{h_{k-1}}-\sqrt{h_k}\right)^2\right]~.
\end{equation}
Here $\mathcal{F}_0$ is a function only of the ratios of the dimensions $h_1,h_2,h_3$.  In this expression, we have kept the leading order singularity as $\theta_k\to 0$, as well as the dominant growth as the $h_k\to\infty$ at fixed but small $\theta$. 
We have neglected logarithmic modifications to the power law, and overall factors that do not scale exponentially with dimension, since they do not contribute to the leading asymptotics, 
just as at genus one. 

Note that in this case, unlike at genus one, we have a constant piece $\mathcal{F}_0^{h_1+h_2+h_3}$ scaling exponentially with dimension. When all dimensions are equal, $h_1=h_2=h_3=h$, we have 
\begin{equation}
\left.\mathcal{F}_0\right|_{h_1=h_2=h_3}=\frac{16}{27} \;
\end{equation}
and when $\sqrt{h_3}=\sqrt{h_1}+\sqrt{h_2}$, $\blockF_0$ is unity. We do not have an explicit expression for $\mathcal{F}_0$ in other cases, but we can calculate it perturbatively in the ratios.  The result is that, for fixed total dimension $h_1+h_2+h_3$, $\mathcal{F}_0$ is minimised when the dimensions are equal. The coefficients $\left(C_{i_1,i_2,i_3}\right)^2$ must have the \emph{opposite} exponential scaling as $	|\mathcal{F}(h_1,h_2,h_3;\theta_k)|^2$ simply to get the correct radius of convergence in \rref{blockSingularity}.  The result is that the three point coefficients will be exponentially larger when the three dimensions $h_1=h_2=h_3$ are equal.

At large dimension, as the density of states becomes large, we may replace the discrete sum over primary operators by a continuous integral over a `density of squared OPE coefficients' at given dimension, $\rho_{C^2}\left(h_k,\bar{h}_k\right)$:
\begin{equation}
	\sum_{i_1,i_2,i_3 \atop \text{primary}} \left(C_{i_1,i_2,i_3}\right)^2 \approx \int d^3 h \; d^3\bar{h} \: \rho_{C^2}(h_k,\bar{h}_k)
\end{equation}
We can then invert \rref{blockSingularity} to determine $\rho_{C^2}$.  To do so, 
it will be convenient to replace the dimensions $h_k$ with `Liouville momenta' $\alpha_k=\sqrt{\frac{6h_k}{c}}$ and conjugates, and replace the moduli $\theta_k$ with (inverse) `temperatures' 
\begin{equation}
\beta_k \equiv \frac{\pi^2}{2\log\frac{1}{\theta_k}}
\end{equation} 
and conjugates, which go to zero in the limit of interest.  Equation \rref{blockSingularity} becomes
\begin{equation}\label{mordor}
	\int d^3\alpha\: d^3\bar{\alpha}\: \rho_{C^2}(\alpha,\bar{\alpha}) e^{-\frac{c}{6}\bm{\alpha}^t\cdot\bm{\beta}\cdot\bm{\alpha}}e^{-\frac{c}{6}\bar{\bm{\alpha}}^t\cdot\bar{\bm{\beta}}\cdot\bar{\bm{\alpha}}} \mathcal{F}_0(\alpha_i/\alpha_j)^{\frac{c}{6}\alpha^2} \bar{\mathcal{F}}_0(\bar{\alpha}_i/\bar{\alpha}_j)^{\frac{c}{6}\bar{\alpha}^2} \approx \exp\left(\frac{c}{24}\pi^2\sum_k\left(\frac{1}{\beta_k}+\frac{1}{\bar{\beta}_k}\right)\right).
\end{equation}
The right hand side here is the power law singularity we want to reproduce, including also the correction to the power law coming from the individual blocks in \rref{blockAsymptotics}. 
We have written the dimensions as a vector $\bm{\alpha}$ of the three `Liouville momenta', and the moduli as a symmetric `temperature matrix' $\bm{\beta}$:
\begin{equation}
\bm{\beta}=
	\begin{pmatrix}
		\beta_1+\beta_2+\beta_3 & \beta_3-\beta_1-\beta_2 & \beta_2-\beta_1-\beta_3 \\
		\beta_3-\beta_1-\beta_2 & \beta_1+\beta_2+\beta_3 & \beta_1-\beta_2-\beta_3 \\
		\beta_2-\beta_1-\beta_3 & \beta_1-\beta_2-\beta_3 & \beta_1+\beta_2+\beta_3
	\end{pmatrix}~.
\end{equation}

We wish to reproduce the right hand side from a small $\beta$, large $\alpha$ saddle point over the integral on the left hand side. To achieve this, $\rho_{C^2}$ must cancel the $\mathcal{F}_0$ piece, and the remainder must scale exponentially with $\alpha$, as
\begin{equation}
	\rho_{C^2}\left(h_k,\bar{h}_k\right) \sim \exp\left[\pi\sum_k\left(\sqrt{\frac{c}{6} h_k}+\sqrt{\frac{c}{6} \bar{h}_k}\right)\right] \mathcal{F}_0^{-(h_1+h_2+h_3)}\bar{\mathcal{F}}_0^{-(\bar{h}_1+\bar{h}_2+\bar{h}_3)},
\end{equation}
a form that factorises into pieces depending only on the separate dimensions\footnote{Putting this $\rho$ into \rref{mordor}, the integral becomes a Gaussian which is sharply peaked at its saddle point.  Noting that \begin{equation}
	(1,1,1)\cdot \bm{\beta}^{-1} \cdot \begin{pmatrix}1\\1\\1\end{pmatrix} = \sum_k\frac{1}{\beta_k},
\end{equation} we find this matches the right hand side, up to subleading prefactors.}. The saddle point is at $\alpha_1=\frac{\pi}{4}(\beta_2^{-1}+\beta_3^{-1})$ and cyclic permutations, so for given dimensions, the surface corresponding to the saddle point has moduli $\theta_1=e^{-\pi(\alpha_2+\alpha_3-\alpha_1)}$. The formula therefore applies only when $\sqrt{h_1}+\sqrt{h_2}\gg\sqrt{h_3}$ (and permutations), so that the saddle point is reproducing the small $\theta$ behaviour.

This is the total contribution to the partition function from all states of dimension $(h_1,h_2,h_3)$.  However, since we already know the asymptotic density of states
, we may divide by density of states to get the average of the square of heavy three point functions.  The result is that
\begin{equation}\label{elros}
	\overline{C_{h_1 h_2 h_3}^2} \sim \exp\left[-\pi\sum_k\left(\sqrt{\frac{c}{6} h_k}+\sqrt{\frac{c}{6} \bar{h}_k}\right)\right] \mathcal{F}_0^{-(h_1+h_2+h_3)}\bar{\mathcal{F}}_0^{-(\bar{h}_1+\bar{h}_2+\bar{h}_3)}~.
\end{equation}
Here the average is taken in the microcanonical sense, over a small window of operator dimensions centred on dimensions $h_k,\bar{h}_k$. When the dimensions are equal, this becomes
\begin{equation}
	\overline{C_{h h h}^2} \sim \left(\frac{27}{16}\right)^{3(h+{\bar h})}\exp\left[-3\pi\left(\sqrt{\frac{c}{6} h}+\sqrt{\frac{c}{6} \bar{h}}\right)\right]~.
\end{equation}
The presence of an exponential term in this expression is not surprising: the three point coefficients are defined in terms of the correlation function with operators located at $0$, $1$ and $\infty$, but this is just a convention. A different choice would change the correlation function by just such an exponential factor.  We could, for example, choose to place the operators on an equilateral triangle of side length $\mathcal{F}_0^{-1}$ ($=27/16$ for all dimensions equal, smaller for unequal dimensions).  In this case there would be no exponential dependence in the three point coefficient.

So far we have worked only at large central charge, where we will (in the next section) be able to use monodromy techniques to determine the asymptotics of the conformal blocks.  We expect that the result at finite central charge $c>1$ will be qualitatively similar.  In particular, we conjecture that the asymptotic behaviour of the blocks at large $h_k$ and finite $c$ takes the same form as \rref{blockAsymptotics}, but possibly with more general coefficients
\begin{equation}\label{blockAsymptoticsGuess}
	\mathcal{F}_{g=2}(h_1,h_2,h_3;\theta_k)\sim \left(\prod_k\theta_k \right)^{-\kappa_1} \mathcal{F}_0^{h_1+h_2+h_3} \exp\left[-\kappa_2\sum_k \frac{\pi^2}{2\log\frac{1}{\theta_k}}\left(\sqrt{h_{k+1}}+\sqrt{h_{k-1}}-\sqrt{h_k}\right)^2\right]
\end{equation}
where the $\kappa_1, \kappa_2, \blockF_0$ are unknown.  The result will be that the three point coefficients will take the same form as \rref{elros} at large dimension
\begin{equation}\label{elrond}
	\overline{C_{h_1 h_2 h_3}^2} \sim \exp\left[\kappa_3\sum_k\left(\sqrt{ h_k}+\sqrt{\bar{h}_k}\right)\right] \mathcal{F}_0^{-(h_1+h_2+h_3)}\bar{\mathcal{F}}_0^{-(\bar{h}_1+\bar{h}_2+\bar{h}_3)}~
\end{equation}
with some unknown coefficients $\kappa_3$ and $\blockF_0$.  The most natural guess is that -- in analogy with the density of states -- when $c>1$ the result is given precisely by \rref{elros}, just with the replacement $c\to c-1$. It would be interesting to investigate this further.

\section{Asymptotics of higher genus blocks}\label{blocks}

In the previous section, we obtained a formula for the asymptotics of OPE coefficients of primary operators. The main technical requirement for the derivation was to account for the contribution of each conformal family to the genus 2 partition function, that is the genus two conformal blocks. In this section we derive those results, justifying the claimed asymptotic formula \cref{blockAsymptotics}.

The main data we require is the form of the blocks when the internal dimensions are large, and at the edge of moduli space corresponding to small $\theta_k$. Understanding this in generality is a hard problem, so we will focus primarily on the blocks in the additional simplifying `semiclassical' limit of large central charge. In this limit, data about the blocks can be obtained by the `monodromy method' due to Zamolodchikov \cite{zamolodchikov1987conformal}, which we will briefly review before explaining how it may be applied to higher genus partition functions. The resulting problem is still not solvable exactly excepting for certain limits and special cases, but when the internal operator dimensions are large, the problem admits a WKB approximation. We use a combination of this WKB method and pinching limit methods applicable for small $\theta_k$ to obtain the desired formulae.

\subsection{The monodromy method}

This subsection is largely a review of the ideas underlying the Zamolodchikov's monodromy method for conformal blocks \cite{zamolodchikov1987conformal}, justified more fully in \cite{Harlow:2011ny}.

Conformal blocks are kinematic objects, depending only on the central charge and dimensions of operators present, so to compute them we may consider any theory we please. In particular, take a CFT of large central charge $c=1+6\left(b+b^{-1}\right)^2$, with $b\ll 1$, which also contains a level 2 degenerate operator $\psi$ of conformal dimension $-\frac{1}{2}-\frac{3}{4}b^2\approx -\frac{1}{2}$. Consider the block of interest (a correlation function, in the case of interest the partition function on some Riemann surface, with operators projecting to some conformal family inserted along some cycles) with an additional insertion of $\psi(z)$. In the semiclassical limit, this is expected to equal the block without the extra insertion, times a function $\psi_c(z)$ which is finite in the large $c$ limit, intuitively the expectation value of $\psi$ in the background created by the other operators. We can now write the condition that the level two null field $(\mathcal{L}_{-1}^2+b^2 \mathcal{L}_{-2})\psi$ decouples as an ODE:
\begin{equation}\label{MonodromyODE}
	\psi_c''(z) + T_c(z) \psi_c(z)=0
\end{equation}
Here, $T_c(z)$ can be understood as $b^2\approx \frac{c}{6}$ times the semiclassical expectation value of the stress tensor, analogously to $\psi_c$.

At this stage, we do not know $T_c(z)$. However, it is highly constrained, since it is holomorphic everywhere except at operator insertions, where by the OPE it has a double pole determined by the operator dimension, and a single pole whose residue do not know \emph{a priori}, but once fixed, gives the derivative of the block by the position of the operator. This unfixed data is therefore very useful: if we know it across moduli space, we can integrate to find the functional form of the block. These residues are not independent, but subject to three relations, interpreted as the Ward identities for the global $\mathfrak{sl}(2)$ part of the conformal symmetry, so this gives $n-3$ parameters. On a Riemann surface of genus $g\geq 1$ there are additional `global' parameters, corresponding to the derivatives of the block with respect to the moduli of the surface, numbering $3g-3$ for $g\geq 2$ or 1 for the torus. Summing up, by these considerations we can specify $T_c$ up to a finite number of `accessory' parameters, which correspond to variations of the block under changes of moduli\footnote{More abstractly, the affine space of fields transforming like a stress tensor is equivalent to the tangent space of the Teichm\"uller space of the Riemann surface with punctures at operator insertions \cite{Friedan:1986ua}.}

So we now have an ansatz for $T_c$, and wish to learn about the block by fixing the remaining free parameters. To do this, we use what we know about the solutions to the ODE \cref{MonodromyODE}, given that we have projected to given conformal families, built on primaries of dimension $h_p$, on some cycles. Near the projection operator, $\psi_c$ will look like a sum of three-point functions of $\psi$, an operator of dimension $h_p$, and an operator to which they fuse. But because $\psi$ is degenerate at level 2, the fusion can only be to two possible dimensions, corresponding to the two independent solutions of the ODE. The two possible three-point functions pick up a factor as as the cycle is traversed, and in terms of the ODE, these factors are the eigenvalues of the monodromy matrix:
\begin{equation}\label{monoEigenvalues}
	\text{Eigenvalues of monodromy} = -e^{\pm i \pi\alpha_p},\quad \text{where } h_p = \frac{c}{24}(1-\alpha_p^2)
\end{equation}
If we impose this monodromy condition on sufficiently many cycles, so that the block is determined kinematically, this will be sufficient to fix all the accessory parameters as desired.

Having fixed the semiclassical stress tensor, we now want to find the block. The change in the block under some variation of the moduli of the Riemann surface can be computed by inserting an appropriately smeared stress tensor into the correlation function, since it corresponds to a variation of the background metric. In the semiclassical limit, a stress tensor insertion just multiplies the correlation function by the semiclassical value $\frac{c}{6}T_c(z)$, which suggests that the block must exponentiate
\begin{equation}
	\mathcal{F}\sim e^{-\frac{c}{6}f}
\end{equation}
with the values of the semiclassical stress tensor giving the derivative of $f$ with respect to the moduli. Denoting the moduli by $x_i$ and the accessory parameters of the semiclassical stress tensor, appropriately defined, by $c_i$, this will give a relationship of the form
\begin{equation}
	\frac{\partial f}{\partial x_i} = c_i.
\end{equation}
We will discuss the precise definition of moduli and accessory parameters needed to make this true in examples.

\subsection{For higher genus blocks}

With the general strategy outlined, let us turn to the case of interest, specifically higher genus blocks (without operator insertions) on a surface specified by an algebraic curve of the form
\begin{equation}
	y^n = \prod_{k=1}^N \frac{z-u_k}{z-v_k}
\end{equation}
which may be alternatively described as a correlation function in the $\ZZ_n$ orbifold of the theory, with $N$ twist operators inserted at the $u_k$ and $N$ anti-twist operators at the $v_k$. We may also take one fewer term in the denominator, which corresponds to taking one of the anti-twist operators to infinity. We start by working out the most general allowed form of the semiclassical stress tensor $T_c$ on such a surface.

Except at the branch points $u_i$ and $v_i$, and at infinity, $z$ is a good coordinate, so $T_c(z)$ is locally analytic in $z$, though may take a different value on each of the $n$ sheets of the cover. At the branch points $u_k$, the $y$ coordinate is regular, so $T_c$ is analytic in that frame instead, which gives a condition in the $z$ frame, that
\begin{equation}
	T_c(z) \sim \frac{1-n^{-2}}{4(z-u_k)^2} + \frac{y^2}{(z-u_k)^2}f_k(y) \quad \text{as }z\to u_k,
\end{equation}
where $f_k(y)$ is a holomorphic function of $y$ in the neighbourhood of $0$ (that is, a power series in $y$ starting at $y^0$). The first term comes from the Schwarzian of the change of coordinates. This form allows for branch cuts like $z^{j/n}$ for integers $j$, so that $T_c(z)$ can take different values on each sheet.

Similarly, at the $v_k$, $y^{-1}$ is a good coordinate, and we have
\begin{equation}
	T_c(z) \sim \frac{1-n^{-2}}{4(z-v_k)^2} + \frac{y^{-2}}{(z-v_k)^2}g_k(y^{-1})\quad \text{as }z\to v_k,
\end{equation}
with $g_k$ holomorphic in $y^{-1}$ at $y=\infty$, and finally, there is a condition for smoothness at $z=\infty$, which demands that $T_c(z)$ decays as $z^{-4}$ or faster. The exception is when there is a branch point at infinity, so there is one fewer factor $z-v_k$ in the denominator than in the numerator, in which case we have
\begin{equation}
	T_c(z) \sim \frac{1-n^{-2}}{4z^2} + \frac{y^{-2}}{z^2}g_\infty(y^{-1})\quad \text{as }z\to \infty.
\end{equation}

These conditions constrain $T_c$ to a $(3g-3)$-dimensional space as required. We may write the most general object on the Riemann surface as
\begin{equation}
	T_c = \sum_{k=1}^n T_c^{(k)}(z)y^k
\end{equation}
where the $T_c^{(k)}(z)$ will turn out to be rational functions of $z$, with poles determined by the above conditions. The $y^k$ allow for functions that are not single value as a function of $z$, but are single-valued on the surface; we need only powers up to $n$ because $y^n$ can be substituted for a rational function of $z$ by the defining relation of the curve.

To show how this works, we will just take two simpler examples. Firstly, consider the case relevant for the $n$th R\'enyi entropy of a pair of intervals, that is $N=2$, putting the branch points at $u_1=0$, $u_2=1$, $v_1=x$, $v_2=\infty$ by a M\"obius map, so the curve is
\begin{equation}
	y^n=\frac{z(z-1)}{z-x}.
\end{equation}
We may write a general ansatz for a stress tensor on the curve as
\begin{align}
T_c(z)&=\frac{1-n^{-2}}{4}\left(\frac{1}{z^2}+\frac{1}{(z-1)^2}+\frac{1}{(z-x)^2}-\frac{2}{z(z-1)}\right)\\
	&+\frac{y^{-1}p_{-1}(z)+p_0(z)+y p_1(z)}{z(z-1)(z-x)}+\frac{1}{z^2(z-1)^2(z-x)}\sum_{k=2}^{n-2}y^k p_k(z)
\end{align}
where the top line is chosen to provide the correct double poles determined by the Schwarzian pieces of the transformation to the $y$ frame (including at infinity), and the second line is a general expression for a holomorphic function on the surface, with the $p_k$ for the moment arbitrary holomorphic functions of $z$. The poles at $0,x,1$ explicitly put in are chosen to be the maximally singular poles allowed by smoothness, so the $p_k$ must also be holomorphic at those points. Smoothness at infinity determines that the function must decay as $z^{-3}y^k$ for $-1\leq k\leq n-2$, which constrains $p_{-1},p_0,p_1$ to be constants, and the remaining $p_k$ to be quadratic polynomials (except when $n=2$, when only the constant $p_0$ term is allowed). In total, this gives $3+3(n-3)=3n-6$ parameters, as expected of a genus $n-1$ surface. The parameter $p_0$ corresponds to deformations that keep the curve in the same form, just deforming the cross-ratio $x$, and retaining the $\ZZ_n$ symmetry.

For the second example, take $n=2$ and $N$ arbitrary. In this case, it is convenient to eliminate the denominator on the right hand side of the algebraic curve, by multiplying by its square and absorbing in the definition of the coordinate on the left hand side. Like this, we may write such a curve as
\begin{equation}
	w^2=f(y)
\end{equation}
where $f$ is a polynomial of degree $2N$, with roots at $y_k$, the insertions of twist operators\footnote{We use the $y$ coordinate here, since it turns out that in some cases, one of which we will use later, it is the same as $y$ in the previous example in an alternative representation of the same surface.}. Such a surface is known as a hyperelliptic curve, as it generalises the $N=2$ case of a torus, also known as an elliptic curve. This surface corresponds to the most general correlation function of twist operators in a $\ZZ_2$ orbifold, reflecting the fact that there is only one twisted sector in that theory (in particular, a twist and anti-twist operator are equivalent at $n=2$).

Now write an ansatz for the stress tensor as
\begin{equation}
	T_c(y)= \frac{3}{16}\sum_{k=1}^{2N}\frac{1}{(y-y_k)^2} + \frac{p(y)+w q(y)}{f(y)}
\end{equation}
where the first term gives the Schwarzian pieces required for regularity at the endpoints, and the denominator in the second term is chosen so that regularity at $y_k$ is equivalent to regularity of $p$ and $q$. These functions are therefore entire analytic, and are further constrained by regularity at infinity, $T_c$ decaying like $y^{-4}$ or faster. This constrains $p$ to be a polynomial of degree $2N-2$, but with the coefficients of the largest powers fixed to cancel the $y^{-2}$ and $y^{-3}$ piece of the Schwarzian term, and $q$ to be a polynomial of degree $N-4$ (since $w\sim y^N$ and $f(y)\sim w^{2N}$). This leaves $3N-6$ parameters, matching the count of moduli of a genus $g=N-1$ curve. The $2N-3$ free parameters in $p$ correspond to the deformations of the surface that leave the curve in the hyperelliptic form, respecting the $\ZZ_2$ symmetry $w\mapsto -w$ (the hyperelliptic involution), which can be achieved by moving the roots $y_k$ (minus the three deformations corresponding to M\"obius transformations of the $y$ plane). 

This sets up the ansatz for the ODE relevant for higher genus blocks. We conclude the subsection by describing how the accessory parameters, once fixed by the monodromy conditions on the ODE, relate to derivatives of the semiclassical block with respect to the moduli of the surface. Specifically, let us focus on deformations that keep the curve in the same form, but shift one of the branch points. It is easiest to consider this operation in the language of the orbifold theory, in which the derivative of the block is implemented by inserting the operator $L_{-1}$ around the twist operator. This operator is the integral of the orbifold theory stress tensor round the circle, which is the integral of the stress tensor of the original theory around the circle $n$ times, passing through all sheets of the cover, to make a closed loop on the Riemann surface. This cancels all pieces of the semiclassical stress tensor that have nontrivial branching, and picks out $n$ times the residue of the remaining piece. If the branch point is at $z_k$
, this gives us
\begin{equation}\label{blockDerivative}
	T_c(z)\sim \sum_{k=0}^{n-1} T_c^{(k)}(z)y^k \implies \frac{\partial f}{\partial z_k} = -n \underset{z= z_k}{\operatorname{Res}} T_c^{(0)}(z)
\end{equation}

\subsection{The large dimension WKB approximation}

The discussion above sets up the calculation of semiclassical higher genus blocks in terms of the monodromy of an ODE. Except for a few special cases (one of which we will see later), this problem is still not exactly solvable, so we have to rely on some approximation. Luckily, in the case of most direct interest for us, when the intermediate dimensions are large, we can apply a WKB approximation to the ODE.

Specifically, let the intermediate dimensions, labelled $h_i$ scale as
\begin{equation}
	h_i\sim \frac{c}{6}\frac{\eta_i}{\lambda^2}
\end{equation}
with $\eta_i$ all independent and fixed, and consider the blocks when $\lambda\ll 1$. We now demand that the eigenvalues of the monodromy matrix are $-e^{\pm \lambda^{-1}2\pi \sqrt{\eta_i}}$, from \cref{monoEigenvalues}. The only way to achieve this is to have the free accessory parameters become asymptotically large in this limit, so we may write
\begin{equation}
	T_c(z)\sim \frac{1}{\lambda^2}T_0(z)+T_1(z)+\cdots,
\end{equation}
which puts the equation in a form where the WKB method is applicable. We then propose an ansatz $\psi_c(z) = e^{\lambda^{-1}\phi(z)}$, with $\phi$ admitting an asymptotic expansion $\phi=\phi_0+\lambda \phi_1+\lambda^2 \phi_2+\cdots$, and solve order by order in $\lambda$. To leading and first subleading order, the independent WKB solutions are
\begin{equation}\label{WKBsolution}
	\phi_0(z)=\pm\int^z \sqrt{-T_0(z)}\; dz,\quad \phi_1(z)=-\frac{1}{4}\log T_0(z),\quad\text{so}\quad \psi(z)\sim \frac{1}{T_0^{1/4}} e^{\pm\lambda^{-1}\int\sqrt{-T_0}}
\end{equation}
which allow us to write down the eigenvalues of the monodromy quite simply, since the monodromy is diagonal in this basis. If $T_0$ is meromorphic inside the cycle of interest, $T_0^{-1/4}$ picks up a factor of $-i$ for each zero and $i$ for each pole of $T_0$ contained within the cycle (counted with multiplicity), and in the cases of interest this will give us $-1$ from a pair of poles. This leaves the remainder of the monodromy to be picked up in the exponential, and we get the result
\begin{equation}\label{monodromyIntegral}
	\eta_k = \left(\frac{1}{2\pi i}\oint_{\gamma_k} \sqrt{T_0(z)}\;dz\right)^2
\end{equation}
where $\gamma_k$ is a cycle on the Riemann surface onto which we project onto the conformal family with weight corresponding to $\eta_k$. Using \cref{monodromyIntegral}, we will be able to fix the order $\lambda^{-2}$ part of the accessory parameters, which will give us a formula for the piece of the block scaling exponentially with dimension.

\paragraph{Stokes phenomena:}

The above considerations are only valid if the two independent WKB solutions remain good approximations to full solutions of the ODE as we go round the contour $\gamma_k$. However, without additional assumptions, this may not be the case, due to Stokes phenomena.  In most places, the two independent WKB solutions remain good approximations to independent solutions of the full ODE, but this is not true everywhere; when certain lines in the complex plane are crossed, a single WKB solution on one of the line may become a linear combination of the two solutions on the other side. These lines are the Stokes lines, where $\Im \phi_0(z)=0$, which typically emanate in threes at angles of $2\pi/3$ from zeros of $T_0$, and divide the complex plane into domains.

If the contour crosses Stokes lines, the integral \cref{monodromyIntegral} will give the incorrect monodromy, and it requires a more careful analysis of the mixing of solutions to proceed. Happily, in all cases we consider, it is possible to choose the contours carefully such that no Stokes lines are crossed, and we will not have to tackle this more involved problem. It is nonetheless necessary to bear in mind, and we will indicate along the way points where na\"ive use of \cref{monodromyIntegral} would fall foul of this subtlety.

\subsection{Pinching limits}\label{pinching}

The asymptotic formula for OPE coefficients uses modular invariance near the edge of moduli space, corresponding to some pinching limit of the Riemann surface, so our main concern will be to understand the block in this limit. As emphasised in \cref{OPEderivation}, to recover the correct asymptotic formula it is not sufficient to know only the exponential scaling of the block with large intermediate dimension; we also require the piece of the block that is independent of $h$, but singular in the pinching limit. This singular piece is visible from the WKB analysis at second order, but since the saddle point we found in the analysis of \cref{OPEderivation} is determined by balancing the contributions of these two terms, they must be the same size, which implies the the WKB analysis must break down (the correction is as large as the leading order). Fortunately, the pinching limit allows for its own approximation to find the semiclassical block, independent from the WKB approximation, but with overlapping validity. Following this through, we will find that the formula we get from the second order WKB analysis is nonetheless still true in the regime we require.


In the twist operator frame we use, the pinching limit occurs when when two of the branch points approach one another, so put the points at $z=0,\epsilon$, with $|\epsilon|\ll 1$. In the case of main interest, the semiclassical stress tensor will look like
\begin{equation}\label{pinchingTc}
	T_c(z)\approx \delta\left(\frac{1}{(z-\epsilon)^2}+\frac{1}{z^2}\right)+\left(\frac{t+\tilde{t}}{z-\epsilon}-\frac{t-\tilde{t}}{z}\right)
\end{equation}
where we have neglected contributions from other operators, which are of order one, and hence negligible, when $z$ is sufficiently small. 
The coefficient of the double poles is determined by the dimension of the twist operators, equal to $\Delta=\frac{nc}{6}\delta$, so $\delta=\frac{1-n^{-2}}{4}$ as determined above, and we will be interested mostly in the case $n=2$, $\delta=3/16$ (though the same analysis here applies similarly for blocks with operators of any dimension inserted at $z=0,\epsilon$, so we will keep $\delta$ general for now). Here $t,\tilde{t}$ are the accessory parameters, related to the semiclassical block by \cref{blockDerivative}
\begin{equation}
	\frac{\partial f}{\partial \epsilon} = -n(t+\tilde{t})\;
\end{equation}
where the factor of $n$ adds up the stress tensor contribution from all sheets, or alternatively converting between the central charge of the seed theory and $\ZZ_n$ orbifold theory.

In the case of interest, we will be looking at the block in a channel with a projection operator inserted between the two branch points, so will be constraining the monodromy of the ODE along a cycle passing between $0$ and $\epsilon$. If we fix the accessory parameters in the limit as $\epsilon\to 0$, the monodromy will blow up, so to keep the monodromy fixed requires them to grow at an appropriate rate. We can constrain their behaviour as $\epsilon\to 0$ by the physical consideration that at a fixed point away from $z=0$, the stress tensor should not diverge as $\epsilon\to 0$. This constrains $\tilde{t}$ to stay finite as $\epsilon\to 0$, but $t$ can diverge as $1/\epsilon$; this means that the $\tilde{t}$ term will be negligible whenever $z\ll 1$ so we may drop it, but the $t$ term is important (this remains true when we are also in the limit of large external dimension relevant to the WKB approximation; $\tilde{t}$ does not get parametrically enhanced relative to $t$).



Considering first the WKB solution, the monodromy integral \cref{monodromyIntegral} is dominated by a contribution from $z$ of order $\epsilon$, as the contour passes between the poles,
\begin{equation}
	\frac{1}{2\pi i}\int \sqrt{T_0} \;dz \supset \frac{1}{2\pi i}\int \sqrt{\frac{\epsilon t_0}{z(z-\epsilon)}}\;dz \sim \frac{\sqrt{t_0\epsilon}}{\pi i}\log\frac{1}{\epsilon}
\end{equation}
where $t_0$ here is the leading order WKB contribution to $t$, so $t\sim \frac{t_0}{\lambda^2} + t_1 +\cdots$. Contributions coming from such pinching poles will dominate all others (though the contour may pass through several such pairs of pinching poles), so we will find $t_0\sim \frac{1}{\epsilon (\log\epsilon)^2}$. The coefficient is determined by considering all monodromies, each cycle fixing a sum of terms like $\frac{\sqrt{t_0\epsilon}}{\pi i}\log\frac{1}{\epsilon}$ to equal $\sqrt{\eta}$, one for each pinching pair of poles the contour passes between, and taking appropriate linear combinations of cycles determines $t_0$ (see \cref{blocksGenus2} for an example).

One can now straightforwardly go beyond this, and construct the WKB solution to second order. We will omit the details, since we will derive the relevant result in an alternative way, but briefly explain the outcome. The second order solution picks up a contribution to the monodromy at order $\lambda$ from passing between the branch points, which like the first order solution is in danger of diverging as we take $\epsilon\to 0$. Cancelling the leading order divergence, required to fix the monodromy at a finite value, specifies that $t_1\sim \frac{1-8\delta}{4\epsilon}$ as $\epsilon\to 0$, giving a $\log\epsilon$ term in the semiclassical block $f$, and hence a power law in the block itself, $\mathcal{F}\sim \epsilon^{-2\Delta +\frac{nc}{24}}$. This gives precisely the behaviour we claimed in \cref{OPEderivation}. However, the second order contribution grows faster than the first order as $\epsilon\to 0$, so when $\epsilon$ is parametrically small relative to $\lambda$, the correction term becomes larger than the leading order WKB solution, and the approximation breaks down. More precisely, the WKB approximation with internal dimensions $h_p$ is valid only when $\frac{h_p}{c}\gg \left(\log\frac{1}{\epsilon}\right)^2$. 
But this breakdown happens precisely in the regime we want to understand the block, since both the leading order WKB and pole from the subleading order piece are important. We therefore must take more care to justify our claim\footnote{Additionally, if $t_0>0$ there may be Stokes phenomena as the contour passes between the poles; the considerations that follow will also bypass this subtlety}.

We therefore now turn to a pinching limit expansion of the blocks that is independent from the WKB approximation (a related method was discussed in \cite{Fitzpatrick:2016mjq}). Consider a situation where all operators approach one another in pairs, so that the semiclassical stress tensor is a sum of terms like $\cref{pinchingTc}$, centred at different points $z_k$, with independent separations $\epsilon_k$ (which we will take to be all small, but of the same order $\epsilon$), and accessory parameters $t_k$, $\tilde{t_k}$ associated to each pair. As already discussed, the accessory parameters $\tilde{t}_k$ will be of order 1 as $\epsilon\to 0$, and $t_k$ of order $1/\epsilon$. We may make different approximations for $T_c$ in different regimes of $z$, with $|z-z_k|\ll 1$ or $|z-z_k|\gg \epsilon_k$, and solve the ODE $\psi_c''(z)+T_c(z)\psi_c(z)=0$ separately in each regime. The regimes of validity of these approximations overlap in the annulus $\epsilon_k \ll |z-z_k|\ll 1$, so we can match the solutions there to compute the monodromy. Here we describe only the salient qualitative features of the solutions that are necessary to derive the small $\epsilon$ behaviour of the blocks, relegating some details to an appendix.

Firstly, when $z$ is not close to any of the operators, so $|z-z_k|\gg \epsilon_k$ for all $k$, we may approximate the pairs of poles by just a double pole at each point $z_k$, contributing
\begin{equation}
	\frac{1-\alpha_k^2}{4(z-z_k)^2} + \frac{2\tilde{t}_k}{z-z_k}\quad \text{with } \frac{1-\alpha_k^2}{4}=2\delta+t_k\epsilon_k
\end{equation}
so as long as we are not too close to any of the $z_k$ we can solve the ODE with half the number of singular points, in terms of the unknown $t_k$, $\tilde{t}_k$. If we are in the region $\epsilon_k \ll |z-z_k|\ll 1$, only the double pole at $z_k$ is relevant, and the solutions are approximately power laws $\psi_c\sim (z-z_k)^{\frac{1\pm\alpha_k}{2}}$. The $\alpha_k$ have a natural physical interpretation: far from $z_k$, the two separate poles are not resolved individually, and appear like the contribution of a single operator, with `effective' dimension parametrised by $\alpha_k$. In the WKB regime, this effective dimension is large and negative (corresponding to large real $\alpha_k$), but it becomes of order one when $\frac{h}{c}$ is of the same order as $(\log\epsilon)^2$, which is the regime of the saddle point in \cref{OPEderivation}.

In the main case of interest for us, there will be three pairs of poles, so the $\tilde{t}_k$ are all fixed by smoothness of $T_c$ at infinity, and since there are three singular points the solutions are given by some hypergeometric ${}_2F_1$ functions. In any case, the salient point is that we may compute the monodromy between the solutions in the regions $\epsilon_k \ll |z-z_k|\ll 1$ along any contour that does not pass between any pair of pinching poles, in the basis of power law solutions $(z-z_k)^{\frac{1\pm\alpha_k}{2}}$ in those regions, and the entries of the monodromy matrix will be of order one (at least if $\alpha_k$ is of order 1; if $\alpha_k$ is large, as in the WKB regime, the analysis remains true, but slightly more care is needed).

Secondly, when $|z-z_k|\ll 1$, we can neglect the poles from other points, as well as the $\tilde{t}_k$ term. It is suggestive to write the resulting $T_c$ as
\begin{equation}
	\epsilon_k^2\;T_c(z)\approx \delta\left(\frac{1}{\left(\frac{z-z_k}{\epsilon_k}-1\right)^2}+\frac{1}{\left(\frac{z-z_k}{\epsilon_k}\right)^2}\right)+\frac{t_k\epsilon_k}{\frac{z-z_k}{\epsilon_k}\left(\frac{z-z_k}{\epsilon_k}-1\right)}
\end{equation}
where all neglected terms vanish in the limit $\epsilon_k\to 0$ for fixed $\frac{z-z_k}{\epsilon_k}$. In this form, it is clear that we can change variables in the ODE to $w=\frac{z-z_k}{\epsilon_k}$, and what remains is an equation for $w$ that depends on $\epsilon$ only through $c_k\epsilon_k$, or alternatively the same $\alpha_k$ as above. Now in the regime $1\ll |w| \ll \epsilon_k^{-1}$, $T_c$ is again well approximated by a double pole, so the solutions are approximately power laws $w^{\frac{1\pm\alpha_k}{2}}$. This matches onto the solutions in the other regime, where $|w|\gg 1$. We may now compute the monodromy matrix along a contour that starts in the annulus $1 \ll |w| \ll \epsilon_k^{-1}$, passes between the poles at $w=0,1$, and goes back to the annulus, and in the $w^{\frac{1\pm\alpha_k}{2}}$ basis the entries will again be order one (with the same comments regarding the size of $\alpha_k$ applying). To match this onto the solution away from the poles, we should change basis to powers of $z$, achieved by conjugating with a diagonal matrix, with entries $\epsilon_k^{(1\pm\alpha_k)/2}$. This results in a monodromy matrix with order one entries on the diagonal, but entries of order $\epsilon_k^{\pm\alpha_k}$ on the off-diagonal.

Now to find the trace of the monodromy around a closed loop, we simply need to string these monodromy matrices obtained from the different approximations together, and take the trace. When $\alpha_k>0$ (which we may choose as long as $\alpha_k^2>0$, which can be verified a posteriori), the term that dominates the matrix product and trace will contain the largest factor $\epsilon_k^{-\alpha_k}$ for every pair of poles it passes through, so the monodromy equation (taking $\alpha_p$ large and imaginary, corresponding to large dimension $\frac{24h_p}{c}\sim -\alpha_p^2\sim 4\eta/\lambda$) is
\begin{equation}
	\Tr M_\gamma = -2\cos(\pi\alpha_p) \implies \mu(\alpha_k) \prod_k \epsilon_k^{-\alpha_k} \sim e^{2\pi\sqrt{\eta}/\lambda}
\end{equation}
for some order one function $\mu$. At leading order, we can neglect $\mu$, and find that $2\pi\sqrt{\eta}/\lambda$ is approximated by a sum of terms $\alpha_k\log\frac{1}{\epsilon_k}$, one for each pair of poles near $z_k$ that the contour passes between; just as in the discussion of the WKB approximation, by fixing the monodromy around a full set of cycles, we can fix the $\alpha_k$, and hence the $t_k$.

Note that this analysis fixes $t_k =- \frac{\alpha_k^2}{4\epsilon_k} + \frac{1-8\delta}{4\epsilon_k} $, and writing these two terms as $\lambda^{-2}t_0$ and $t_1$, we find that $t_1$ matches the value obtained from the second order WKB solution, and $t_0$ obeys precisely the same condition in relation to the monodromies as the leading order WKB integral. The conclusion is that the blocks indeed have the claimed asymptotic behaviour, even in this regime beyond the validity of WKB.

We close this discussion with a brief summary of the conditions for the pinching limit analysis presented to be valid; much of this is discussed more fully in the appendix. The approximations we have made require that $\alpha_k\log\frac{1}{\epsilon_k}$ is large, $\alpha_k$ is not too close to zero, and  we also require that the prefactors like $\mu$ do not become large or small to interfere with the separations of scales we have used. When $\alpha_k$ is large, as in the case for the WKB approximation, it is in principle possible that the coefficients of the other terms in the monodromy trace that we have neglected could become large enough to compensate for their suppression, but in fact this does not happen, so the regime of validity overlaps with the WKB regime. This can be understood as resulting from a simplification of the monodromy matrices at large $\alpha$, because in the first regime (away from poles) they become diagonal and in the second regime (going between the poles) they become off diagonal in the power law basis, since it is also the WKB basis. The approximation could also break down if the prefactor $\mu$ approaches a zero, interfering with the dominance of this term. This in fact happens at some order one value of $\alpha$ if $\delta<3/16$, so in those cases there is a crossover to new behaviour for sufficiently small $\epsilon$. In the case of interest, where $\delta=3/16$, $\mu$ remains positive for all positive values of $\alpha$, so this does not happen, as verified in the appendix. In addition the analysis as described is not quite true when some $\alpha_k$ is parametrically close an integer, since the monodromy matrices are not all order one, but receive an additional logarithmic enhancement; this only affects subleading behaviour, so does not alter the conclusions.

Finally, we note that the saddle point, as we will see in \cref{blocksGenus2}, corresponds to the values $\alpha_k$ all equal to one, which is far from the small $\alpha_k$ regime where the analysis breaks down. This has a very natural interpretation: the saddle point is chosen to reproduce a pole in the cross channel coming from the identity operator, so the `apparent dimension', corresponding to the strength of the double pole in $T_c$ as seen far from the pinching twist operators, should be zero.

\subsection{Warmup: the blocks at genus one}\label{blocksGenus1}

At genus one, the algebraic curve description of a torus is as an elliptic curve, which may be brought to the form
\begin{equation}
	y^2=z(z-1)(z-x)
\end{equation}
for some $x$; this is the description of the torus partition function as the four-point function of twist operators in a $\ZZ_2$ orbifold theory \cite{Lunin:2000yv,Headrick:2010zt}. We know the blocks of interest exactly, since they are just Virasoro characters for the torus partition function, written in the appropriate coordinates and conformal frame, but as a warm-up exercise for the genus two case, let us nonetheless approach the problem in the semiclassical WKB approximation.

The most general ansatz for the semiclassical stress tensor now has one free parameter $t$:
\begin{equation}
	T_c(z)=\frac{3}{16}\left(\frac{1}{z^2}+\frac{1}{(z-1)^2}+\frac{1}{(z-x)^2}-\frac{2}{z(z-1)}\right)
	+\frac{t}{z(z-1)(z-x)}
\end{equation}
Note that this contains no freedom to have a stress tensor value that is different on the two sheets, so the semiclassical genus one block will be identical to the semiclassical Virasoro block in the orbifold theory, of central charge $2c$, with external operators of dimension $c/16$, and with an exchange operator of dimension $2h$. It should be emphasised that they are different for finite $c$, since the genus one block contains contributions from extra states; this equivalence simply says that these extra states are negligible at large $c$.

Following the WKB approximation described above, let us write $t=\lambda^{-2}t_0+\cdots$, so $T_0=\frac{t_0}{z(z-1)(z-x)}$, and impose a monodromy condition on a cycle surrounding $0$ and $x$, corresponding to projecting onto the conformal family of a dimension $h=\frac{c\eta}{6\lambda^2}$ primary:
\begin{equation}
	\sqrt{\eta} =\sqrt{t_0}\frac{1}{2\pi i}\oint_{\gamma} \frac{dz}{\sqrt{z(z-1)(z-x)}} = \sqrt{t_0}\frac{1}{\pi i}\int_0^x \frac{dz}{\sqrt{z(x-z)(1-z)}} = \sqrt{t_0}\frac{2}{\pi i} K(x)
\end{equation}
The last integral gives us the elliptic $K$ function, which determines $t_0$.

We now use \cref{blockDerivative} to relate the accessory parameter to the derivative of the block:
\begin{equation}
	\frac{\partial f}{\partial x} =\frac{2t}{x(1-x)}
\end{equation}
To find the block to leading order in the WKB approximation we now need only to integrate $t_0(x)$
\begin{equation}
	f(x)\sim -\lambda^{-2}\eta\int \frac{\pi^2}{2x(1-x)K(x)^2}dx = \frac{6h}{c}\; 2\pi\frac{K(1-x)}{K(x)}+\text{constant}
\end{equation}
which gives the result of Zamolodchikov for the large $h$ asymptotics of the Virasoro block:
\begin{equation}\label{genus1WKB}
	\mathcal{F}_{g=1}\approx e^{-\frac{c}{6}f} \approx \exp\left(-2h\pi\frac{K(1-x)}{K(x)}\right) =\exp\left(2h\pi i\tau\right)=q^h
\end{equation}
Here $\tau=i \frac{K(1-x)}{K(x)}$ is the usual modular parameter on the torus, and $q=e^{2\pi i\tau}$ is the nome, and the normalisation is fixed to match the conventions of \cref{CardyDerivation}, with the twist operator OPE coefficients absorbed into the block so that each primary operator in the seed theory contributes precisely $\mathcal{F}\bar{\mathcal{F}}$ to the twist operator correlation function.

In fact it turns out that the solution \cref{WKBsolution} from the WKB approximation is an \emph{exact} solution to the ODE, with accessory parameter $t=\frac{1-2x}{8}+\frac{\pi^2}{16K(x)^2}\left(1-\frac{24h}{c}\right)$, which allows us to find the semiclassical block without using the further approximation of large dimension \cite{Maloney:2016kee}. This gives the result quoted in \cref{CardyDerivation}, only missing the eta function contribution from the descendants, which is invisible at large $c$:
\begin{equation}
	\mathcal{F}^\text{(semiclassical)}_{g=1} = (2^8x(1-x))^{-c/24} q^{h-c/24}
\end{equation}

If we now look at the $x\to 1$ limit of the block, we find a power law singularity $(1-x)^{-c/24}$. This comes from the limit of the accessory parameter $t\to \frac{1}{8}$ as $x\to 1$, which matches the expectation from the analysis above, with a single pinching pair of poles (though note that the accessory parameter here differs from the convention of the previous section by a factor of $-x(1-x)$). We also reproduce the correct $x\to 1$ behaviour $\exp\left[h\frac{2\pi^2}{\log(1-x)}\right]$ of the leading order WKB contribution, with the coefficient fixed by $2\pi\lambda^{-1}\sqrt{\eta}=\alpha\log\frac{1}{1-x}$, with $\alpha$ as defined in that section, and only the piece of the contour passing between the poles at $z=x,1$ contributing to the monodromy at leading order.

\subsection{The blocks at genus two}\label{blocksGenus2}

There are two nice frames to work in at genus 2, corresponding to the threefold cover of the Riemann sphere branched at four points, or the twofold cover branched at six points. We will work mainly with the latter representation, since it is more general, allowing for the full three parameter moduli space to be mapped out, but when working with the special one-parameter family of $\ZZ_3$ symmetric surfaces described by the threefold cover, we will at times find it convenient to use that parameterisation. For the symmetric cases, the surfaces may be written as algebraic curves
\begin{equation}
	y^3=\frac{z(z-1)}{(z-x)},\quad \text{equivalent to }  w^2 = \left(y^3-e^{i\theta}\right)\left(y^3-e^{-i\theta}\right)\quad\text{with } x=\cos^2\frac{\theta}{2}
\end{equation}
so the branch points in the $y$ frame are located at the roots $u_k = e^{i(2\pi k+\theta)/3},v_k = e^{i(2\pi k-\theta)/3}$ of $\left(y^3-e^{i\theta}\right)\left(y^3-e^{-i\theta}\right)$, collectively denoted by $y_l$. We will refer to these two descriptions as the $z$-frame and the $y$-frame, referring to the coordinate in which we choose the standard flat metric on the complex plane in the two cases. We will be considering the partition function in the limit $x\to 1$, corresponding to $\theta\to 0$. For the leading order WKB, large $h$ piece of the block, we can be lazy about specifying which frame we work in, since this limit is insensitive to the contribution of the conformal anomaly.

Referring to our earlier results for the most general allowed ansatz for $T_c$, we find
\begin{equation}
	T_c(z)=\frac{2}{9}\left(\frac{1}{z^2}+\frac{1}{(z-1)^2}+\frac{1}{(z-x)^2}-\frac{2}{z(z-1)}\right)+\frac{p(y)}{y z(z-1)(z-x)}
\end{equation}
in the $z$ frame, where $p$ is a quadratic in $y$, and in the $y$ frame, this becomes
\begin{equation}
	T_c(y)= \frac{3}{16}\left(\sum_{m=1}^{6}\frac{1}{(y-y_m)^2}-\frac{6y^4}{w^2}\right) + \frac{\tilde{p}(y)}{w^2}
\end{equation}
where $\tilde{p}$ is quadratic in $y$. In terms of the $z$-frame parameters, we have $\tilde{p}(y)=9 p(y)+\frac{7}{4}y \cos\theta$.

The block of interest requires projecting onto a conformal family of a primary of dimension $h_1,h_2,h_3$ around three independent cycles $\gamma_1,\gamma_2,\gamma_3$, around which the monodromy condition will be imposed. In the $z$ frame, these cycles surround the branch points at $0$ and $x$, with one for each sheet of the cover. In the $y$ frame, they each surround a pair of the branch points; choose $\gamma_k$ so that it surrounds the branch points at $-e^{2\pi k i/3\pm i(\pi-\theta)/3}$.
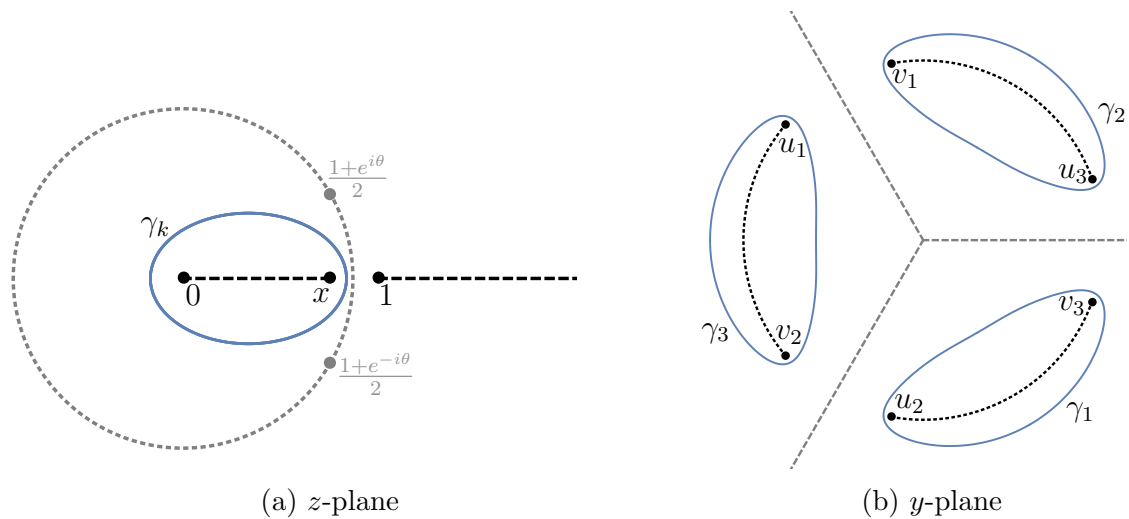
\begin{figure}
	\centering
	\begin{subfigure}[b]{.5\textwidth}
	\begin{picture}(100,100)
		\put(0,0){\includegraphics[width=.9\textwidth]{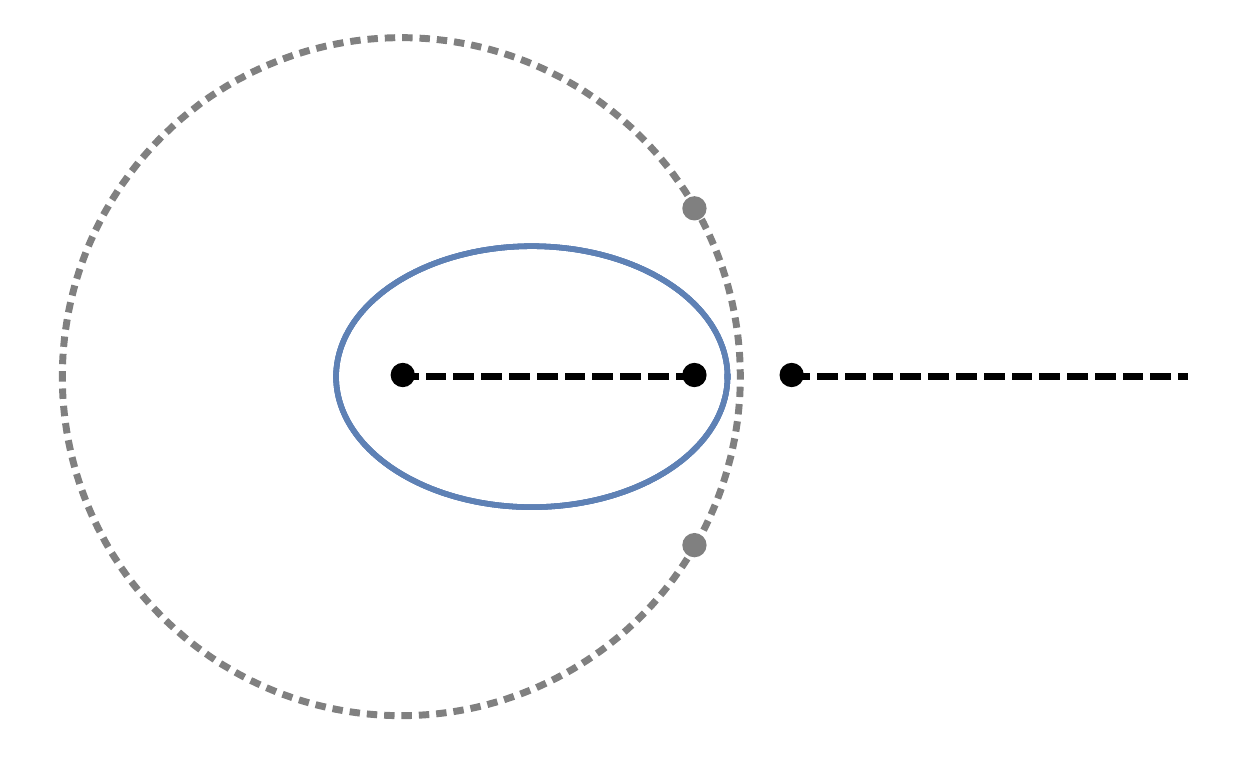}}
		\put(60,90){$\gamma_k$}
		\put(77,62){$0$}
		\put(125,63){$x$}
		\put(150,62){$1$}
		\put(130,107){\textcolor{Gray}{$\frac{1+e^{i\theta}}{2}$}}
		\put(134,32){\textcolor{Gray}{$\frac{1+e^{-i\theta}}{2}$}}
	\end{picture}
	\caption{$z$-plane}
	\end{subfigure}
	\begin{subfigure}[b]{.35\textwidth}
	\begin{picture}(200,200)
		\put(0,0){\includegraphics[width=.9\textwidth]{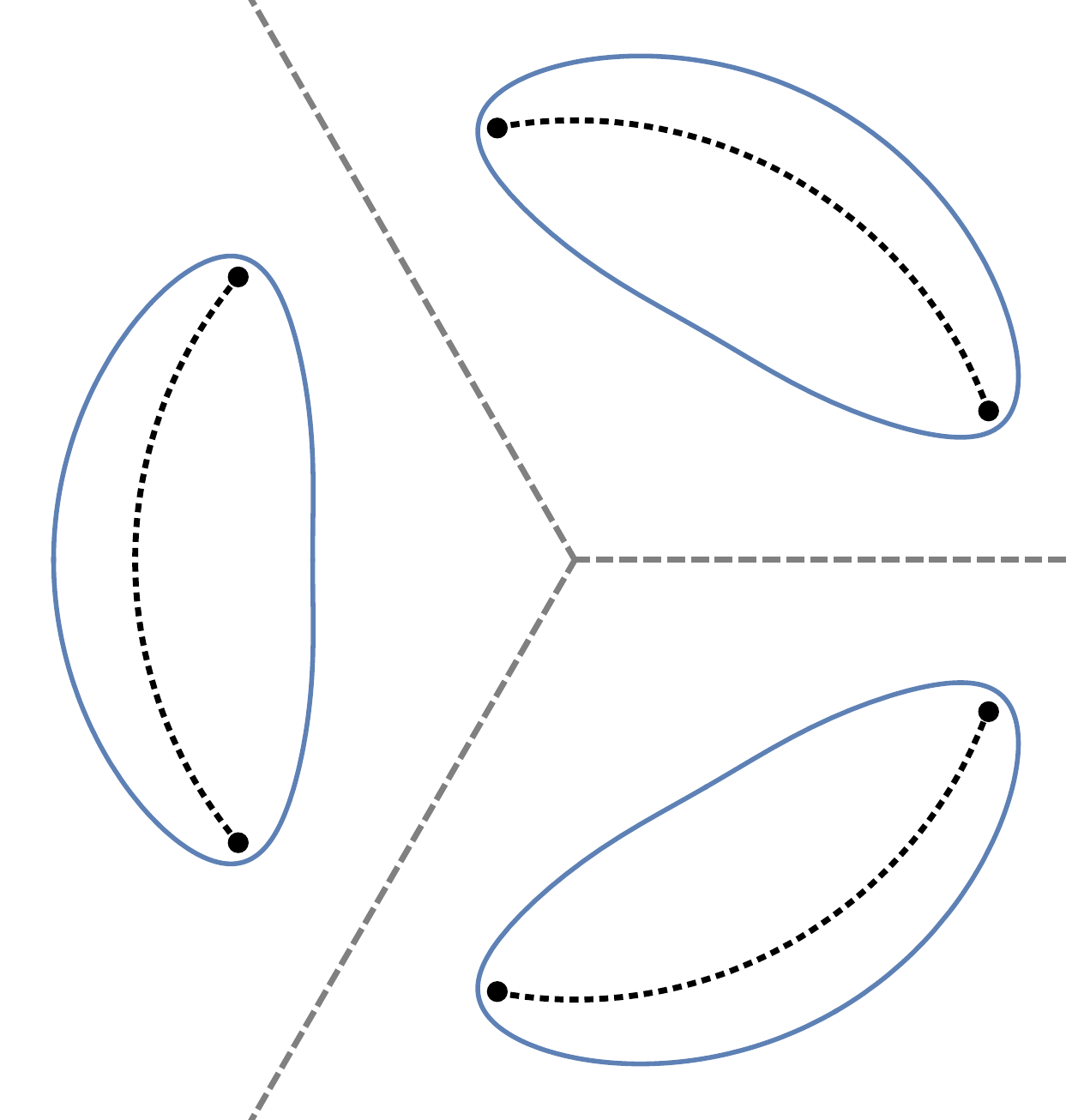}}
		\put(143,20){$\gamma_1$}
		\put(155,133){$\gamma_2$}
		\put(5,50){$\gamma_3$}
		\put(34,120){$u_1$}
		\put(77,146){$v_1$}
		\put(78,23){$u_2$}
		\put(33,49){$v_2$}
		\put(140,110){$u_3$}
		\put(140,60){$v_3$}
	\end{picture}
	\caption{$y$-plane}
	\end{subfigure}
	\caption{Branch points and cuts, and cycles $\gamma_k$ in the $z$ and $y$ planes. Dashed lines are branch cuts on the $z$ plane, and dotted lines branch cuts on the $y$ plane. The blue loops are the cycles $\gamma_k$ around which we fix the monodromy. The two sheets of the $y$ plane correspond to the inside and outside of the dotted circle in the $z$ plane, with the points $z=\frac{1}{2}(1+e^{\pm i\theta})$ mapping to $y=u_k,v_k$, and the three sheets of the $z$ plane correspond to the three wedges of the $y$ plane separated by the dashed lines.\label{zyFig}}
\end{figure}

Using the $y$ frame, we may straightforwardly generalise away from the $\ZZ_3$ symmetric case, by putting the roots at $u_k=e^{i\frac{2\pi k+\theta_k}{3}}$, $v_k=e^{i\frac{2\pi k-\theta_k}{3}}$, with independent $\theta_k$, $k=1,2,3$. We may use the results of \cref{pinching} to find the relevant asymptotics of the block, up to an overall normalising constant. We have three pairs of pinching points $(u_k,v_k)$ near $e^{2\pi i k/3}$, with separations of approximately $\epsilon_k=2\theta_k/3$ (up to unimportant phases). With accessory parameters $t_k$ chosen so that $\frac{\partial f}{\partial \epsilon_k}\approx-2t_k$, we have $t_k \approx -\frac{1}{8\epsilon_k}-\frac{\alpha_k^2}{4\epsilon_k}$, with the $\alpha_k$ fixed by the dimensions of the internal operators, $2\pi\lambda^{-1}\sqrt{\eta_k}=\alpha_{k-1}\log\frac{1}{\epsilon_{k-1}}+\alpha_{k+1}\log\frac{1}{\epsilon_{k+1}}$, because the contour $\gamma_k$ passes between two pairs of poles, at $e^{2\pi i (k\pm1)/3}$. Taking the sum of two of these equations, and subtracting the third, we isolate one of the accessory parameters, getting $\alpha_k^2=\frac{(2\pi)^2}{\lambda^2(\log\epsilon_k)^2}(\sqrt{\eta_{k+1}}+\sqrt{\eta_{k-1}}-\sqrt{\eta_{k}})$. It is then straightforward to put this all together, integrate and exponentiate, to get the block asymptotic formula (keeping approximation only to the relevant order)
\begin{equation}
	\mathcal{F}_{g=2}\approx \mathcal{F}_0^{h_1+h_2+h_3} \left(\prod_k\theta_k \right)^{-\frac{c}{24}}  \exp\left[-\sum_k \frac{\pi^2}{2\log\frac{1}{\theta_k}}\left(\sqrt{h_{k+1}}+\sqrt{h_{k-1}}-\sqrt{h_k}\right)^2\right]
\end{equation}
as claimed in \cref{blockAsymptotics}.

All that remains is to fix the normalisation, coming from the integration constant for the semiclassical block. The only relevant part of the normalisation for our main result comes from the leading order WKB block, scaling exponentially with internal dimensions, written as $\mathcal{F}_0$, which depends only on the ratios $h_i/h_j$ of dimensions, kept fixed in the heavy limit. The normalisation of the block is conventional, but once we have fixed the normalisation in the $\theta_k\to \pi$ limit, which determines the coefficient with which the blocks appear in the OPE decomposition of the correlation function, the factor as $\theta_k\to 0$ is determined, so the relative normalisation is meaningful, captured by $\mathcal{F}_0$.

To calculate it requires knowing the accessory parameters along a family of moduli joining the $\theta_k\to 0$ and $\theta_k\to \pi$ limits, so that we may integrate along the whole family. A simple choice that achieves this is the family of enhanced $\ZZ_3$ symmetry, with all the $\theta_k$ equal, alternatively parameterised by the cross-ratio $x$ in the three-sheeted frame. We will focus for the remainder of the section on the WKB calculation for these curves, moving freely between the two frames as convenient; key landmarks in the two descriptions are marked in \cref{zyFig} for orientation.

For the na\"ive WKB calculation, we must compute the integrals
\begin{equation}\label{genus2WKBints}
	I_k = \frac{1}{2\pi i}\oint_{\gamma_k} \sqrt{T_0(z)}\;dz = \frac{1}{2\pi i}\oint_{\gamma_k} \sqrt{\frac{t_- y^{-1}+t_0 +t_+ y}{z(z-1)(z-x)}}\;dz =\frac{3}{2\pi i}\oint_{\gamma_k} \sqrt{\frac{t_-+t_0 y+ t_+ y^2}{(1+y^3)^2-4x y^3}}\;dy = \sqrt{\eta_k}
\end{equation}
where we have written the integral in both frames, and $y$ in the $z$ representation is defined implicitly by $y^3=\frac{z(z-1)}{z-x}$, the branch of the solution depending on $k$. We then need only to solve $I_k=\sqrt{\eta_k}$ for the accessory parameters $t_0,t_\pm$ in terms of $\eta_k$, and $t_0$ determines the derivative of the block with respect to $x$ as in \cref{blockDerivative}:
\begin{equation}
	\frac{\partial f}{\partial x} \sim \frac{3 t_0}{\lambda^2 x(1-x)}
\end{equation}
The accessory parameters $t_\pm$ do not appear in this formula, since they are dual to deformations that break the $\ZZ_3$ symmetry, orthogonal to deformation by varying $x$. Equivalently, they do not appear in the semiclassical expectation value of the orbifold stress tensor, which is $c/6$ times the sum of $T_c$ on the three sheets, and the phases from $y^{\pm 1}$ cancel in the sum.

Comparing with the genus one calculation, we might be tempted to take $\gamma_k$ to be the cycle in the $z$ frame going along the straight line from $0$ to $x$, and back again on the other side of the branch cut. But this is tricky, because $T_0(z)$ has zeros, coming from the quadratic in $y$ in the numerator, and these lead to additional branch cuts in the integrand, and associated Stokes phenomena in the WKB approximation. In the $y$ frame, this choice of contour goes from the origin, to infinity along a ray with argument $\frac{(2k-1)\pi}{3}$, and back from infinity to zero along a ray with argument $\frac{(2k+1)\pi}{3}$. At least one of these contours is bound to contain one or both zeros of $T_0$, which guarantees that it will cross Stokes lines. We will therefore choose alternative contours to avoid this, which contain one of the branch cuts in the $y$ plane from the denominator of $T_0$ as they must, but do not contain either of the zeros of the numerator of $T_0$.

Nonetheless, these integrals are hard to do in general, so we must make some approximation to make progress. One way of computing $I_k$ is to expand the integrand as a series in $x$, so that at each term in the expansion, the branch cut from the numerator disappears, and the integrand has only poles inside the contours. Then, term by term, $I_k$ can be computed by the residue at $-e^{2\pi i k/3}$. This gets more and more difficult as the order in $x$ increases, but is an efficient way to compute for a few orders in the $x$ expansion. To the first two orders in $x$, we get
\begin{equation*}
	\mkern-24mu I_k^2 =(e^{-2\pi i k/3}t_-+e^{2\pi i k/3}t_+-t_0) + \left(\frac{4(e^{-2\pi i k/3}t_-+e^{2\pi i k/3}t_+-t_0)}{9}+\frac{t_0^2-4 t_- t_+}{18(e^{-2\pi i k/3}t_-+e^{2\pi i k/3}t_+-t_0)}\right)x+O(x^2)
\end{equation*}
and it is then straightforward to solve $I_k^2=\eta_k$ order by order for $t_0,t_+,t_-$, linear equations at each order in $x$, with the following result for $t_0$:
\begin{equation}
	t_0 = -\frac{1}{3}(\eta_1+\eta_2+\eta_3) + \left[\frac{\eta_1+\eta_2+\eta_3}{6} -\frac{1}{162}\left(\frac{(\eta_2-\eta_3)^2}{\eta_1}+\frac{(\eta_3-\eta_1)^2}{\eta_2}+\frac{(\eta_1-\eta_2)^2}{\eta_3}\right)\right]x +\cdots
\end{equation}
Integrating up to find the large $h$ block, we get
\begin{equation*}
	\mkern-18mu \log\mathcal{F}_{g=2} \underset{h\to \infty}{\sim}  (h_1+h_2+h_3)\log x + \left[\frac{h_1+h_2+h_3}{2}+\frac{1}{54}\left(\frac{(h_2-h_3)^2}{h_1}+\frac{(h_3-h_1)^2}{h_2}+\frac{(h_1-h_2)^2}{h_3}\right)\right]x +\cdots
\end{equation*}
where we have not yet fixed the constant of integration to normalise.

While this may be useful information about the blocks for other purposes, it does not give us what we need, since in practice we can only compute to some finite order in $x$. While we are not able to find an exact result for all $x$ in generality, we can do it in the case when the dimensions are equal, or approximately equal by perturbing in the ratio of the dimensions. To do this, note first that for all dimensions equal, the block is $\ZZ_3$ symmetric, so we expect $t_-=t_+=0$. In this situation, the calculation becomes the same as the usual Virasoro block in the $z$ frame, and the asymptotics are the same as those worked out by Zamolodchikov \cite{Zamolodchikov:1985ie} as reviewed in \cref{blocksGenus1}. Going beyond that case, we may take $t_\pm$ to be nonzero but small, and expand the numerator of $T_0$ in a series in these small parameters. Having done that, the terms in the expansion look like
\begin{equation}
	\frac{3}{2\pi i}\oint_{\gamma_k} \frac{y^{p+1/2}}{\sqrt{(1+y^3)^2-4x y^3}}\;dy = i\left(-e^{2\pi i k/3}\right)^p {}_2F_1\left(\frac{1}{2}-\frac{p}{3},\frac{1}{2}+\frac{p}{3};1;x\right)
\end{equation}
for positive and negative integers $p$. The integral can be done by the same method of expanding in $x$ and evaluating residues as above, but now it is simple enough to do at all orders in $x$ to get a hypergeometric series. Write $\eta_k = 1 + \delta\eta_k$, where we have normalised so the average $(\eta_1+\eta_2+\eta_3)/3$ is fixed at 1, so $\delta\eta_1+\delta\eta_2+\delta\eta_3=0$; we may now solve the monodromy equation for $t_0,t_\pm$, order by order in $\delta\eta$.

If we define the convenient shorthand
\begin{equation}
	F_a(x) := {}_2F_1\left(a,1-a;1;x\right) = F_{1-a}(x)
\end{equation}
for the hypergeometric functions appearing, the result for $t_0$ to the first couple of orders is 
\begin{align}
	& t_0 =-F_\frac{1}{2}^{-2}+\frac{1}{6}\left(\delta\eta_1^2+\delta\eta_1 \delta\eta_2 +\delta\eta_2^2\right)\left(F_\frac{1}{2}^{-2}-F_\frac{1}{6}^{-2} \right)  \\& \qquad -\frac{\delta\eta_1 \delta\eta_2\delta\eta_3}{8}\left(F_{\frac{1}{2}}^{-2}-F_{\frac{1}{6}}^{-2}-F_{-\frac{1}{2}}F_{\frac{1}{6}}^{-3}+F_{\frac{1}{2}}F_{-\frac{1}{6}}F_{\frac{1}{6}}^{-4}\right) +O(\delta\eta^4) \nonumber
\end{align}
where it should be noted in particular that there is no linear term in $\delta\eta$, given the choice $\delta\eta_1+\delta\eta_2+\delta\eta_3=0$.
To quadratic order in $\delta\eta$, we can integrate the block explicitly by introducing new $\tau$ parameters, and noting a formula for their derivative\footnote{This follows from writing the equation as the Wronskian of the hypergeometric differential equation, of which $F_{1/n}(x)$ and $F_{1/n}(1-x)$ are independent solutions.}, generalising the relations for the usual $\tau$ parameter above (which is the case $n=2$):
\begin{equation}
	\tau_n(x) := i \frac{F_{1/n}(1-x)}{F_{1/n}(x)},\quad \tau_n'(x) = \frac{\sin(\pi/n)}{i\pi}\frac{1}{x(1-x)F_{1/n}(x)^2}
\end{equation}
The usual $\tau$ parameter is $\tau_2$ here. We will use the asymptotics for small $x$, given by
\begin{equation}
	i\pi\tau_n(x) \sim \sin\left(\frac{\pi}{n}\right)\log\left(\frac{x}{a_n}\right) + O(x)
\end{equation}
for some constants $a_n$ (useful values are $a_2=2^4,a_3=3^3,a_4=2^6,a_6=2^43^3$), to fix the constant term in the blocks, from \cref{smallTheta}. Translating to the $x$ variable via $\pi-\theta\sim 2\sqrt{x}$ as $x\to 0$, with our normalisation the blocks behave as
\begin{equation}
	\mathcal{F}_{g=2}\sim 
	\left(\frac{x}{27}\right)^{h_1+h_2+h_3}\left(\frac{16x}{9}\right)^{-3c/16} (1+O(x))
\end{equation}

Integrating the result for $t_0$, and fixing the constant as $x\to 0$, gives us
\begin{align}
	&\qquad\lim_{h\to\infty} \frac{\log \mathcal{F}_{g=2}}{h_1+h_2+h_3} = -\int \frac{t_0}{x(1-x)} dx\\ 
	=& (i\pi \tau_2+4\log 2-3\log3) +\frac{1}{6}\left(\delta\eta_1^2+\delta\eta_1 \delta\eta_2 +\delta\eta_2^2\right)\left(i\pi \left(2\tau_6 - \tau_2\right)+3\log 3\right) +O(\delta r^3) \nonumber
\end{align}
where the relevant $x\to 0$ behaviour of this expression is $\log x-3\log 3+O(x)$. It is a useful check that the logarithmic term in the perturbation cancels; it also matches the small $x$ expansion derived above in the appropriate regime.

From this, we may read off the constant $\mathcal{F}_0$, simply by taking $x\to 1$ and noting that $\tau_n\to0$ in this limit:
\begin{equation}
	\mathcal{F}_0 = \frac{16}{27}\left(1+\frac{1}{2}\log 3 \left(\delta\eta_1^2+\delta\eta_1 \delta\eta_2 +\delta\eta_2^2\right)+\cdots \right)
\end{equation}
Note in particular that the correction term is positive definite, which will mean that for fixed average dimension, perturbing away from equal dimensions will exponentially suppress the OPE coefficients at large dimension.

We can also find the heavy asymptotic block for all values of $x$ at the edge of the regime of validity of our asymptotic formula, when $\sqrt{h_1}+\sqrt{h_2}=\sqrt{h_3}$ (or some permutation). To do this, notice that the WKB integrals \cref{genus2WKBints} simplify if $t_-+t_0 y+t_+ y^2$ is a perfect square, when $t_0^2=4t_- t_+$, evaluating to a multiple of $F_{1/3}(x)$ by the same methods as used before. If we parametrise the accessory parameters as
\begin{equation}
	t_-+t_0y+t_p y^2=\left(\frac{e^{-i\pi/6}(\sqrt{\eta_1}y-\sqrt{\eta_2})+e^{i\pi/6}(\sqrt{\eta_2}y-\sqrt{\eta_1})}{\sqrt{3}F_{1/3}(x)}\right)^2
\end{equation}
we find that the integrals evaluate to $I_k^2=\eta_k$ as required by the monodromy condition, with $\eta_3=\left(\sqrt{\eta_1}+\sqrt{\eta_2}\right)^2$. To find the large $h$ blocks, we expand the square to get the linear term in $y$, $t_0=-\frac{\eta_1+\eta_2+\eta_3}{3 F_{1/3}(x)^2}$, which we can integrate as before:
\begin{equation}
	\lim_{h\to\infty} \frac{\log \mathcal{F}_{g=2}}{h_1+h_2+h_3} = \frac{2\pi i}{\sqrt{3}}\tau_3 
	\implies \blockF_{g=2} \sim q_3^\frac{h_1+h_2+h_3}{2},\quad\text{where}\quad q_3=e^{\frac{4\pi i}{\sqrt{3}}\tau_3}
\end{equation}
Note in particular that we have not needed to add any constant term to fix the correct normalisation. Since $\tau_3\to 0$ as $x\to 1$, in particular this means that $\blockF_0=1$ when $\sqrt{h_1}+\sqrt{h_2}=\sqrt{h_3}$, so the exponential term is absent from our asymptotic formula at its boundary of validity. This is closely analogous to the formula \cref{genus1WKB} for the genus one blocks.

\section*{Acknowledgements}

We are grateful to A. Belin, M. Cho, S. Collier, J. Kaplan, C. Keller, P. Kraus, and X.Yin for useful conversations.  
J.C. is supported in part by funds from the Simons Foundation.
A.M. and H.M. are supported by the National Science and Engineering Council of Canada and by the Simons Foundation.  

\appendix
\section*{Pinching limits of genus 2 blocks}\label{appendix}

In this appendix, we give some details of the calculation of the semiclassical genus two conformal blocks in the pinching limit as described in \cref{pinching}. We will do this in the frame where the blocks are given by the six-point function of twist operators in a $\ZZ_2$ orbifold theory, which is equivalent to the usual semiclassical Virasoro block for the six point function of dimension $c/32$ operators. It will be convenient to do this in a different frame, where the pairs of operators are meeting near $0$, $1$ and $\infty$.

Here, we will calculate in detail the monodromy matrix round one of the poles near $z=0$ and one of the poles near $z=1$. This is sensitive to the separations $\epsilon_{0,1}$ of the pairs of poles near $z=0,1$.

When we are not close to the branch points, $T_c(z)$ is well approximated by
\begin{equation}
	T_c(z)\approx\frac{z^2-z+1-\alpha_0^2 (1-z)-\alpha_1^2 z+\alpha_\infty^2 (1-z) z}{4 (z-1)^2 z^2}
\end{equation}
where we parametrise the unknown accessory parameters by the coefficients $\frac{1-\alpha_k^2}{4}$ of double poles at $z=0,1,\infty$, with $k=0,1,\infty$ respectively. In this region, the ODE is solved by hypergeometric functions:
\begin{equation}
	\psi_{\pm}(z) = z^\frac{1\pm\alpha_0}{2}(1-z)^\frac{1+\alpha_1}{2} {}_2F_1\left(\frac{1\pm\alpha_0+\alpha_1-\alpha_\infty}{2},\frac{1\pm\alpha_0+\alpha_1+\alpha_\infty}{2};1\pm\alpha_0;z\right)
\end{equation}
This basis of solutions is chosen to approximate a power $z^\frac{1\pm\alpha_0}{2}$ when $z$ is small. We can use an alternative basis of solutions which become powers $(1-z)^\frac{1\pm\alpha_1}{2}$ near $z=1$, by exchanging $z\leftrightarrow 1-z$ and $\alpha_0\leftrightarrow \alpha_1$. By standard hypergeometric identities, the change of basis matrix is
\begin{equation}
M_{10}=
	\begin{pmatrix}
		\frac{\Gamma(1+\alpha_0)\Gamma(-\alpha_1)}{\Gamma\left(\frac{1+\alpha_0-\alpha_1-\alpha_\infty}{2}\right)\Gamma\left(\frac{1+\alpha_0-\alpha_1+\alpha_\infty}{2}\right)} & \frac{\Gamma(1-\alpha_0)\Gamma(-\alpha_1)}{\Gamma\left(\frac{1-\alpha_0-\alpha_1-\alpha_\infty}{2}\right)\Gamma\left(\frac{1-\alpha_0-\alpha_1+\alpha_\infty}{2}\right)} \\
		\frac{\Gamma(1+\alpha_0)\Gamma(\alpha_1)}{\Gamma\left(\frac{1+\alpha_0+\alpha_1-\alpha_\infty}{2}\right)\Gamma\left(\frac{1+\alpha_0+\alpha_1+\alpha_\infty}{2}\right)}& \frac{\Gamma(1-\alpha_0)\Gamma(\alpha_1)}{\Gamma\left(\frac{1-\alpha_0+\alpha_1-\alpha_\infty}{2}\right)\Gamma\left(\frac{1-\alpha_0+\alpha_1+\alpha_\infty}{2}\right)}
	\end{pmatrix}
\end{equation}
with rows and columns interchanged by swapping the signs of $\alpha_0$ or $\alpha_1$. This is also the monodromy matrix for solutions of the ODE between the regions of small $z$ and small $1-z$, in the local power law bases (the notation is chosen so that $M_{10}$ takes a solution in a power law basis near $0$, and takes it to a solution in the power law basis near $1$). The inverse $M_{01}=M_{10}^{-1}$ is of the same form, with $\alpha_0$ and $\alpha_1$ exchanged.

Looking now near $z=0$, we can change variables to $w=z/\epsilon_0$, and take the $\epsilon\to 0$ limit at fixed $w$, so the ODE becomes
\begin{equation}
	\psi_c''(w)+\left[\frac{3}{16}\left(\frac{1}{w^2}+\frac{1}{(w-1)^2}\right)+\frac{1+2\alpha_0^2}{8w(1-w)}\right] \psi_c(w)=0\; .
\end{equation}
This has solutions
\begin{equation}
	\psi_c(z) = (w(w-1))^{1/4} \left(\frac{\sqrt{w}+\sqrt{w-1}}{2}\right)^{\pm\alpha_0}
\end{equation}
which become power laws when $w$ is large. Note that these are the large $\alpha_0$ WKB solutions, which turn out to be exact in this special case. From these solutions, it is easy to work out the monodromy matrix on passing between the poles at $z=0,\epsilon_0$, after which the sign of one of the square roots is flipped, and a phase comes from the prefactor. It is purely off-diagonal, so it swaps the two solutions in this basis, up to a factor. The simple off-diagonal form is special to this dimension of external operators, and follows from WKB exactness, since WKB solutions do not mix (unless there are Stokes phenomena). After conjugating by a diagonal matrix with entries $\epsilon_0^{\pm\alpha_0}$, which changes to the $z^\frac{1\pm\alpha_0}{2}$ basis from the $w^\frac{1\pm\alpha_0}{2}$ basis, we have the monodromy matrix
\begin{equation}
	M_0 = \begin{pmatrix}
		0& i\left(\frac{4}{\epsilon_0}\right)^{\alpha_0} \\ i\left(\frac{\epsilon_0}{4}\right)^{\alpha_0}& 0
	\end{pmatrix}.
\end{equation}
The monodromy matrix for passing between the poles at $z=1,1-\epsilon_1$ in the $(1-z)^\frac{1\pm\alpha_1}{2}$ basis is the same (make a similar change of variables to $w=(1-z)/\epsilon_1$), up to relabelling $\epsilon_0$ and $\alpha_0$ to $\epsilon_1$ and $\alpha_1$.

We now have the ingredients to find the trace of the monodromy around a cycle going from $0$ to $1$, round a pole near $1$, back to $0$, and round a pole near $0$ to form a closed enclosing the two poles,
\begin{equation}
	\Tr M=\Tr M_0 M_{01} M_1 M_{10} = \left(\frac{4}{\epsilon_0}\right)^{\alpha_0} \left(\frac{4}{\epsilon_1}\right)^{\alpha_1} \mu(\alpha_0,\alpha_1,\alpha_\infty) + \text{(3 terms)},
\end{equation}
where the three additional terms are obtained by changing the sign of $\alpha_0$, $\alpha_1$, or both, and the coefficients are given by
\begin{equation}
	\mu(\alpha_0,\alpha_1,\alpha_\infty)= \frac{\Gamma(\alpha_0)\Gamma(1+\alpha_0)\Gamma(\alpha_1)\Gamma(1+\alpha_1)}{\Gamma\left(\frac{1+\alpha_0+\alpha_1-\alpha_\infty}{2}\right)^2\Gamma\left(\frac{1+\alpha_0+\alpha_1+\alpha_\infty}{2}\right)^2}\; .
\end{equation}

Because of the factor $\epsilon_0^{-\alpha_0}\epsilon_1^{-\alpha_1}$, with the $\alpha_k$ positive, the term that is written explicitly is expected to dominate over the other three, and the results in the text follow from this assumption. However, this domination can be prevented by two possible problems: the coefficient of the dominant term becomes small, or the coefficient of a subdominant term becomes large enough for that term to compete.

The first possibility arises from zeros of $\mu$, which give rise to a crossover to different qualitative behaviour. At sufficiently large internal dimensions, when the $\alpha_k$ are large, we are far from any zeros, and in the regime of interest. Reducing the internal dimensions reduces the $\alpha_k$, and if they reduce to within order $\epsilon$ of a zero of $\mu$, the monodromy becomes dominated by the zero, and the $\alpha_k$ cross over to a power law behaviour approaching the zero. Alternatively, if no zeros are encountered when the $\alpha$ are all positive, the blocks will eventually cross over to a new behaviour dictated by small $\alpha$, where the pole of $\mu$ at $\alpha=0$ becomes important, and the other terms may no longer be suppressed.

In any case, we will stay in the regime of interest as long as the $\alpha_k$ are larger than zero, and remain the correct side of all zeros of $\mu$. The zeros come from poles of the $\Gamma$-functions in the denominator, and the important zero occurs when $\alpha_1+\alpha_0-\alpha_\infty+1=0$, so as long as we remain in the region where $\alpha_1+\alpha_0+1>\alpha_\infty$ and all $\alpha_k$ are positive, the block behaves as we have claimed. This is satisfied at the saddle point, where all the $\alpha$ are equal to one.

The second possible problem, with neglected terms becoming large, na\"ively appears to occur because $\mu$ has poles at negative integral $\alpha_{0,1}$. In fact, this is a problem with the analysis of the monodromy breaking down, and does not affect the blocks: with a more careful analysis, the poles are resolved to logarithmic in $\epsilon$ enhancements, not strong enough to compete with the power law suppression.

Recall that the hypergeometric function is defined by a power series
\begin{equation}
	{}_2F_1(a,b;c;z)=\sum_{n=0}^{\infty} \frac{a(a+1)\cdots(a+n-1)b(b+1)\cdots(b+n-1)}{c(c+1)\cdots(c+n-1)}z^n
\end{equation}
which blows up when $c$ approaches a nonpositive integer. Looking at the solution $\psi_-$ to the ODE away from $0,1,\infty$, given by a hypergeometric function with $c=1-\alpha_0$, we find that it is undefined when $\alpha_0$ is a positive integer $n$, because the terms for $z^n$ and higher have poles at $\alpha_0=n$. Also, when $\alpha_0$ is nearly but not exactly integral, with $\alpha_0=n+\delta$ ($\delta\ll1$), the series for $\psi_-$, with leading order small $z$ behaviour $z^\frac{1-\alpha_0}{2}$, also contains a term like $\frac{1}{\delta}z^{\frac{1+\alpha_0}{2}-\delta}$. For small $\delta$, this will resemble a power law $z^\frac{1+\alpha_0}{2}$, which will contaminate the $\psi_+$ solution when we change to the power law basis.

Because of this, when $\delta$ is sufficiently small, there is a nontrivial change of basis matrix between the hypergeometric basis, and the $w^\frac{1\pm\alpha_0}{2}$ power law basis, with an off-diagonal component:
\begin{equation}
	\begin{pmatrix}
		\epsilon^\frac{1+\alpha_0}{2} & k \,\epsilon^{\frac{1+\alpha_0}{2}}\frac{1}{\delta}\epsilon^{-\delta} \\
		0 & \epsilon^\frac{1-\alpha_0}{2}
	\end{pmatrix},\quad \text{with } k= \frac{(-1)^n \Gamma(1+\delta)}{\Gamma(\alpha_0)}\frac{\Gamma\left(\frac{1-\alpha_0+\alpha_1-\alpha_\infty}{2}+n\right)\Gamma\left(\frac{1-\alpha_0+\alpha_1+\alpha_\infty}{2}+n\right)}{\Gamma\left(\frac{1-\alpha_0+\alpha_1-\alpha_\infty}{2}\right)\Gamma\left(\frac{1-\alpha_0+\alpha_1+\alpha_\infty}{2}\right)}
\end{equation}
The coefficient $k$ stays finite as $\delta\to 0$.

Including the off-diagonal piece, we may recompute the monodromy and get additional terms. The main effect is to resolve the pole in $\mu(\alpha_0,\alpha_1,\alpha_\infty)$ at $\alpha_0=n$. Roughly speaking, it replaces the double pole as follows:
\begin{equation}
	\frac{1}{\delta^2}\longrightarrow \left(\frac{1-\epsilon_0^{-\delta}}{\delta}\right)^2
\end{equation}
When $\delta\gg\log\epsilon$, the correction is not important, but when $\delta$ becomes parametrically small, it replaces the pole in the coefficient of $\epsilon_0^{\alpha_0}\epsilon_1^{-\alpha_1}$ (for example): when $\delta\ll\log\epsilon$, the coefficient is merely enhanced by a factor of $(\log\epsilon_0)^2$. The leading order terms we have used to find the behaviour of the block, going like $\epsilon_0^{-\alpha_0}\epsilon_1^{-\alpha_1}$, are not affected, and the subleading terms remain smaller.

\bibliographystyle{ssg}
\bibliography{biblio}

\end{document}